\documentclass[journal=jacsat,manuscript=article]{achemso}
\usepackage[LGR,T1]{fontenc}
\usepackage[utf8]{inputenc}
\usepackage{color}
\usepackage{array}
\usepackage{textcomp}
\usepackage{multirow}
\usepackage{graphicx}
\usepackage{subscript}
\usepackage{verbatim}
\usepackage{soul}

\makeatletter

\DeclareRobustCommand{\greektext}{%
  \fontencoding{LGR}\selectfont\def\encodingdefault{LGR}}
\DeclareRobustCommand{\textgreek}[1]{\leavevmode{\greektext #1}}
\ProvideTextCommand{\~}{LGR}[1]{\char126#1}

\usepackage{chemformula}
\author{Iker Ortiz de Luzuriaga}
\affiliation[nanogune]
{CICnanoGUNE BRTA, Tolosa Hiribidea 76, E-20018, Donostia - San Sebastian.}
\alsoaffiliation[UPV]
{Polimero eta Material Aurreratuak: Fisika, Kimika eta Teknologia, Kimika Fakultatea, Euskal Herriko Uniberstitatea, UPV/EHU, 20080 Donostia, Euskadi, Spain.}
\author{Sawssen Elleuchi}
\affiliation[sfax]
{Laboratoire de Chimie Inorganique, LR17ES07, Université de Sfax, Faculté de Sciences de Sfax, 3000 Sfax, Tunisia.}
\author{Emilio Artacho}
\affiliation[CAMB]
{Theory of Condensed Matter, Cavendish Laboratory, University of Cambridge, J. J. Thomson Ave., Cambridge CB3 0HE, United Kingdom}
\alsoaffiliation[ikerbasque]
{Ikerbasque, Basque Foundation for Science, 48011 Bilbao, Spain}
\alsoaffiliation[nanogune]
{CICnanoGUNE BRTA, Tolosa Hiribidea 76, E-20018, Donostia - San Sebastian.}
\alsoaffiliation[DIPC]
{Donostia International Physics Center, 20018 Donostia, Spain.}
\author{Khaled Jarraya}
\affiliation[sfax]
{Laboratoire de Chimie Inorganique, LR17ES07, Université de Sfax, Faculté de Sciences de Sfax, 3000 Sfax, Tunisia.}
\author{Xabier Lopez}
\email{xabier.lopez@ehu.es}
\affiliation[DIPC]
{Donostia International Physics Center, 20018 Donostia, Spain.}
\alsoaffiliation[UPV]
{Polimero eta Material Aurreratuak: Fisika, Kimika eta Teknologia, Kimika Fakultatea, Euskal Herriko Uniberstitatea, UPV/EHU, 20080 Donostia, Euskadi, Spain.}
\author{Adrià Gil}
\email{adria.gil.mestres@csic.es; adriagilmestres@unizar.es; agmestres@fr.ul.pt}
\affiliation[nanogune]
{CICnanoGUNE BRTA, Tolosa Hiribidea 76, E-20018, Donostia - San Sebastian.}
\alsoaffiliation[ARAID]
{ARAID Foundation, Zaragoza, Spain}
\alsoaffiliation[CSIC]
{Departamento de Química Inorgánica, Instituto de Síntesis Química y Catálisis Homogénea (ISQCH) CSIC- Universidad de Zaragoza, C/ Pedro Cerbuna 12, 50009, Zaragoza, Spain}
\alsoaffiliation[BioISI]
{BioISI – Biosystems and Integrative Sciences Institute, Faculdade de Ciências, Universidade de Lisboa, Campo Grande,  1749-016, Lisboa, Portugal}

\title[An \textsf{achemso} demo]
  {Semi-empirical and Linear-Scaling DFT Methods to Characterize duplex DNA and G-quadruplexes in Presence of Interacting Small Molecules}

\makeatother

\begin{document}
\begin{abstract}
The computational study of DNA and its interaction with ligands is a highly relevant area of research, with significant consequences for developing new therapeutic strategies. However, the computational description of such large and complex systems requires considering interactions of different types simultaneously in a balanced way, such as non-covalent weak interactions (namely hydrogen bonds and stacking), metal-ligand interactions, polarization and charge transfer effects. All these considerations imply a real challenge for computational chemistry.  The possibility of studying large biological systems using quantum methods for the entire
system requires significant computational resources, with improvements in parallelization and optimization of theoretical
strategies. Computational methods, such as Linear-Scaling Density Functional Theory and DLPNO-CCSD(T), may allow performing \emph{ab initio} QM calculations, including explicitly the electronic structure for large biological systems, 
at a reasonable computing time. In this work, we study the interaction
of small molecules and cations with DNA (both duplex-DNA and G-quadruplexes), comparing
different computational methods: a linear-scaling DFT (LS-DFT) at
LMKLL/DZDP level of theory, semi-empirical methods (PM6-DH2 and PM7), mixed QM/MM, and DLPNO-CCSD(T). Our goal is to demonstrate the adequacy of LS-DFT to treat the different types of interactions present in DNA-dependent systems. We show that LMKLL/DZDP using SIESTA can yield very accurate geometries and energetics in all the different systems considered in this work: duplex DNA (dDNA), phenanthroline intercalating dDNA, G-quadruplexes, and Metal-G-tetrads considering alkaline metals of different sizes.  As far as we know, this is the first time that full G-quadruplex geometry optimizations have been carried out using a DFT method thanks to its linear-scaling capabilities. Moreover, we show that LS-DFT provides high-quality structures, and some semi-empirical Hamiltonian can also yield suitable geometries. However, DLPNO-CCSD(T) and LS-DFT are the only methods that accurately describe interaction energies for all the systems considered in our study.

\end{abstract}

\section{Introduction}

The interaction of ligands with DNA is a vital research subject with critical therapeutic consequences. Small molecules have shown enough antitumoral activity\cite{neidle2011cancer} binding efficiently to duplex DNA (dDNA) or G-quadruplexes (GQ). For instance, cis-platin\cite{rosenberg1969platinum} is an effective drug which effectively binds to dDNA. However, the toxicity,\cite{braun2011balancing,chua2011clinical} resistance and nonspecific interactions of available drugs make desirable the quest for new molecules that target DNA, in its canonical dDNA form or in other secondary DNA structures, such as G-quadruplexes. In this sense, theoretical methods that help elucidate the nature of the interactions between ligands and DNA are of paramount importance.  Studies on the interaction of ligands with dDNA structures are available in the literature.\cite{denny2001dna,baraldi2004dna,nelson2007non,hamilton2012natural} However, studies with other DNA structures, such as GQ, are scarcer.\cite{arola2008stabilisation,zhang2012recent} GQ DNA structures are formed by the stacking of G-tetrads, each one composed by the planar arrangement of four guanine bases.\cite{balasubramanian2011targeting} GQ have raised considerable interest during the past years for the development of therapies against cancer. These non-canonical structures of DNA may be found in telomeres and/or oncogene promoters, and it has been observed that the stabilization of such GQ may disturb tumor cell growth.\cite{burger2005g,mikami2008antitumor}

There are different theoretical methods available for the study of DNA and the state-of-the-art for the GQ modeling has been reviewed very recently in our team.\cite{ortiz2021learning} These methods range from MD simulations with classical force fields to accurate quantum mechanics (QM) calculations to understand specific local interactions in detail. The use of classical force fields, such as OL15 or bsc,\cite{zgarbova2015refinement,ivani2016parmbsc1}
for the study of DNA has demonstrated high precision and reliability\cite{galindo2016assessing} but they show limitations to deal with GQ since their parameters are usually optimized considering dDNA structures and not GQ. 
In this sense, optimization of force field (FF) parameters for the correct description
of the GQ has been the subject of work by Sponer et al.\cite{zgarbova2015refinement} 

Despite the complexity of biological systems, many phenomena may be studied just taking into account a relatively localized region.  Assuming that the QM treatment is needed, the choice is between a QM cluster approach or the use of the hybrid QM/MM approach.\cite{banavs2009theoretical} Although the QM/MM approach has been extensively used to date,\cite{lin2007qm,magalhaes2020modelling} it also presents shortcomings that may affect the quality of the results: \emph{a})  The QM/MM method is based on partitioning the system of interest in a QM and a MM region, which introduces some arbitrariness in the calculation.  \emph{b}) The partition can lead to the cleavage of covalent bonds, and one needs an appropriate treatment of this boundary. \emph{c}) Electrostatic interactions between the QM and MM regions may be considered at different levels. 
If one needs to overcome the inherent limitations of an artificial QM/MM partition or the property of interest cannot be localized in a small region, a full semi-empirical (SE) description of the system may be a good alternative.\cite{hostavs2013performance,christensen2016semiempirical} SE methods fill the gap between MM and first-principles QM methods, being few orders of magnitude slower than MM methods but still orders of magnitude faster than first-principles QM methods. On the other hand, SE methods may provide reasonable accuracy in geometries and energies, although this is often system-dependent. In addition, SE methods can take polarization and charge transfer
effects into account but they have usually problems describing dispersion and hydrogen bonds.\cite{dannenberg1997hydrogen} Nevertheless, more recently, corrections have been added to overcome these difficulties to improve their performance considerably.
\cite{thiriot2009combining,faver2011formal,yang2008description,tirado2008performance,korth2011empirical} That is, corrections were added by Rezac et al. in 2009 in one of the most popular Hamiltonians, PM6, to improve the description of non-covalent interactions in the modified PM6-DH Hamiltonian,\cite{rezac2009semiempirical} which later was revised in the PM6-DH2 Hamiltonian by Korth et al.\cite{korth2010transferable,yilmazer2013comparison,yilmazer2015benchmark}
Recently, in 2016, an extensive review of the SE methods for non-covalent
biochemical interactions have been carried out by Qiang Cui et al.\cite{christensen2016semiempirical}

The software improvements on theoretical methods and their optimization for parallelization make possible nowadays accurate
\emph{ab initio} calculations, in which the computational cost scales linearly with the system size, such as LS-DFT\cite{bowler2012methods} or near linear-scaling Coupled-Cluster.\cite{riplinger2013efficient} The treatment of the entire system with a first-principles quantum method ensures an accurate description of
the interaction between ligands and biomolecules. Therefore, it can be of high relevance to study DNA-Ligand interactions. 

In the present paper, we use a LS-DFT approach, namely, with the 
SIESTA\cite{soler2002siesta} code, using the LMKLL density functional,\cite{lee2010higher} and the DZDP basis set,\cite{lee2010higher} to analyze the interaction of small chemical species with dDNA and GQ. The results are compared with the semi-empirical methods PM6-DH2\cite{rezac2009semiempirical,korth2010transferable}
and PM7\cite{stewart2013optimization}  (using the MOPAC\cite{james2016stewart} package), QM/MM calculations, near linear-scaling DLPNO-CCSD(T) calculations\cite{riplinger2013natural}
(implemented in the ORCA \emph{v4.2.1} package\cite{neese2012orca}), and benchmark database.\cite{saitow2017new} Modeling of GQ have been the subject of substantial interest recently\cite{ortiz2021learning}. As far as we know, this is the first time that LS-DFT methods have been used to study ligand-GQ interactions by taking into account the whole GQ structure of more than 1000 atoms without reduction to smaller models. We also demonstrate the adequacy of these LS-DFT approaches to treat DNA in its multiple forms and their interactions with small chemical species.

\section{Methods}

To perform SE geometry optimizations, we used MOPAC \emph{v8.0.0} with
the PM6-DH2 and PM7 Hamiltonians. The default Eigenvector Following routine
(EF) was used,\cite{baker1986algorithm} whereas the default SCF criterion
was changed to 1x10\textsuperscript{-8}. The minimum trust radius was
set at 0.0001 $\textrm{Å}$/rad and a damping factor of 10 was added with the SHIFT method to improve the SCF procedure.\cite{https://doi.org/10.1002/jcc.540090203} For the LS-DFT calculations, SIESTA \emph{4.1-b3} software was used.\cite{soler2002siesta}
Geometry optimizations were performed with the LMKLL van der Waals functional,\cite{lee2010higher} which includes dispersion corrections, being highly appropriate for the characterization of weak forces within the ligand-DNA interaction.
The modified Broyden algorithm was used for geometry optimization.\cite{johnson1988modified} SCF convergence was accelerated with the Pulay method\cite{banerjee2016periodic} keeping a history of 4 past density matrices, the density matrix mixing weight was
set to 0.005. For the basis set, a 30 meV energy shift was used along with a 150 Ry mesh cut off for real space integration.
We also used a SCF tolerance of 1x10\textsuperscript{-5} eV and we set the max force tolerance at
0.02 eV/$\textrm{Å}$ for the dDNA and at 0.1 eV/$\textrm{Å}$ in the case of GQ structures. In the case of GQ systems the max force tolerance was established after doing some tests where it was observed that although the tolerance is quite high, it does not compromise either the total energy or the relaxed geometry, at the same time that it allows to reduce the computation time (see Table S6).
An optimized double zeta plus double polarization (DZDP)
basis set was used for each atom\cite{junquera2001numerical} along with Troullier-Martins norm-conserving pseudopotentials\cite{troullier1991efficient, kleinman1982efficacious} that were generated with the
ATOM package included in SIESTA software, whereas for the G-tetrads used for energetic calibration,
the psml pseudopotentials\cite{garcia2018psml} from the website www.pseudo-dojo.org
were used. It must be mentioned that, in the SIESTA method the matrix elements are computed with linear-scaling algorithms while the diagonalization is proportional to \emph{O}(N\textsuperscript{3}).

DLPNO-CCSD(T)\cite{riplinger2013natural} single-point
calculations were also performed with the ORCA \emph{4.2.1} software.\cite{neese2012orca}
The Ahlrichs's def2-SVP basis set\cite{weigend2005balanced} was
used for all the calculations with the corresponding auxiliary bases of Weigend for RI-J\cite{weigend2006accurate}
and RIJCOSX approximations.\cite{kossmann2009comparison} 

We also run QM/MM geometry optimizations for the 1n37 and
2jwq PDB structures, that were performed at M11L/6-31+G(d,p):AMBER
and B3LYP-D3(GD3BJ)/6-31+G(d,p):AMBER level, respectively, as implemented in Gaussian\emph{16}.\cite{g16}
Quadratic Convergence (QC)\cite{bacskay1981quadratically} SCF procedure
was used with a maximum amount of 1500 cycles and the maximum size
for an optimization step was changed from 30 to 1. In both cases the
containing ligand was treated as the QM part and the rest of the molecule
(the DNA) as the MM layer, no boundaries needed to be used, because the ligand
and DNA are not linked.

\subsection{Considered structures }

\subsubsection{DNA base pairs. }
Initial geometries were taken from the benchmark
database performed by Hobza and coworkers, which contains computationally
optimized and experimental DNA base pair structures
interacting through H-bonds and by \(\pi-\pi\) stacking.\cite{jurevcka2006benchmark}
Interaction energies ($\Delta E_{int}$) for these systems were
computed by single-point calculations, at the different levels we compare in this study, and subtracting to
the total energy ($E_{tot}$) the energy of each DNA base fragment
($E_{frag}$) as shown in Eq \ref{eq: DNA_Eint}:

\begin{equation}
\Delta E_{int}=E_{tot}-E_{frag1}-E_{frag2}\label{eq: DNA_Eint}
\end{equation}

\subsubsection{Intercalated Phenanthroline (phen) in DNA base pairs. }

Optimized structures at M06-2X/6-31+G(d,p) level were taken from our previous work\cite{gil2015intercalation} and single-point calculations were performed at the different levels of calculation we compare in this work. The interaction energy for the phen/DNA systems was calculated as shown in Eq \ref{eq: phen/DNA_Eint}, where one fragment is composed of all DNA atoms and the second is the phenanthroline ligand:

\begin{equation}
\Delta E_{int}=E_{tot}-E_{DNA}-E_{phen}\label{eq: phen/DNA_Eint}
\end{equation}

\subsubsection{G-tetrads. }

We considered four model systems from the work of Fonseca Guerra
and coworkers: G\textsubscript{4}MG\textsubscript{4}, aG\textsubscript{4}MG\textsubscript{4},
GQM, and GQ\textsubscript{4Na}M, where M corresponds to the different metal cations (Li, Na, K, Rb, and Cs) placed in the ion channel\cite{zaccaria2016role}
The first two models (G\textsubscript{4}MG\textsubscript{4} and aG\textsubscript{4}MG\textsubscript{4}) only contain the metal cation and the guanine
bases, whereas the other two models (GQM and GQ\textsubscript{4Na}M) contain also the sugar and phosphate backbone or side loop,
which is terminated by H\textsuperscript{+} or Na\textsuperscript{+}, respectively, 
to compensate for the negative charge of the phosphate group.\cite{nieuwland2020understanding}
For these systems, single-point calculations were performed at the different levels of calculation we compare in this work and the interaction energy was calculated following
the same formula as in the original work, Eq \ref{eqn:Celia-eq}:

\begin{equation}
\Delta E{}_{int}=E\left(G_{4}MG_{4}\right)_{gas}-E\left(G_{4}G_{4}\right)_{gas}-E\left(M\right)_{gas}\label{eqn:Celia-eq}
\end{equation}

where G\textsubscript{4}MG\textsubscript{4} is replaced by aG\textsubscript{4}MG\textsubscript{4},
GQM, or GQ\textsubscript{4Na}M, and the empty
scaffold \\G\textsubscript{4}G\textsubscript{4} is replaced
by aG\textsubscript{4}G\textsubscript{4}, GQ, or GQ\textsubscript{4Na}, which are the same structures without the metal cation.

\subsubsection{1n37 and 2jwq systems}
To perform larger geometry optimizations, two different DNA structures
were taken from the PDB: 1) the 1n37 octamer\cite{searle2003dna}, which contains Respinomycin D ligand intercalated in a dDNA
structure; and 2) the GQ structure 2jwq,\cite{hounsou2007g} which has
two MMQ-1 units, each bounded at the end-staking of both sides of
the GQ. The systems were neutralized by adding an alkaline cation close to each phosphate group at a distance of 2.8 $\textrm{Å}$, Na\textsuperscript{+} in the dDNA and K\textsuperscript{+} for the GQ. In the case of the GQ, two additional K\textsuperscript{+} were added, centered
in the ion-channel between G-tetrads, resulting in a +2 charged system.

\section{Results and Discussion}

\subsection{Geometrical discussion}

\subsubsection{Duplex DNA}
\begin{figure}
\includegraphics[scale=0.7]{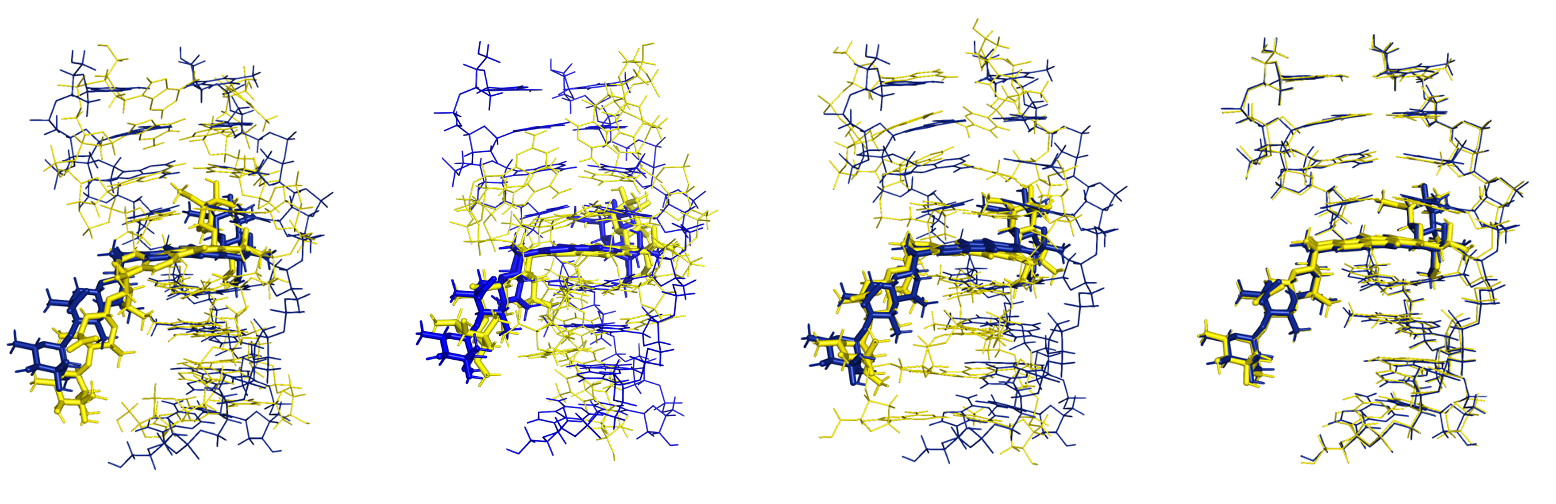}\caption{Overlap structures of 1n37 (in blue) with the optimized geometries
(in yellow). From left to right with the methods PM6-DH2,PM7, M11L/6-31+G(d,p):AMBER and LMKLL/DZDP. RMSD values of 1.84 $\textrm{Å}$, 3.02 $\textrm{Å}$, 1.78 $\textrm{Å}$ and 0.45 $\textrm{Å}$ were obtained, respectively.}
\label{fgr:duplexDNA_structures} 
\end{figure}

Geometry optimizations were run departing from the PDB structures
with SE, QM/MM, and LS-DFT methods. In Figure \ref{fgr:duplexDNA_structures},
we plot the superposition of the 1n37 structure from the PDB with the optimized
geometries. SE results are the ones that shows the highest deviation from the 
initial reference structure giving the highest Root-Mean-Square Deviation (RMSD) values. Namely, 
PM6-DH2 shows a RMSD of 1.84 $\textrm{Å}$, although the structure seems to be maintained. However, PM7 gives a much higher RMSD, 3.02
$\textrm{Å}$, and the general structure is clearly not well superimposed. QM/MM method gives a slightly better result than PM6-DH2 with a 1.78 $\textrm{Å}$ RMSD value and a lower deviation for the structure corresponding to the ligand than for the dDNA part. Finally, the LS-DFT method is by far the best method reproducing the PDB structure, giving an excellent RMSD value of 0.45 $\textrm{Å}$.
Such results show that the SIESTA method and software not only has
good performance in terms of computing time for large systems of thousands of atoms, but also that the obtained structures are very accurate in terms of geometry. 

To further calibrate the performance of these methods for structural characterization, we also used other geometrical parameters characteristic of these systems with drugs intercalating between
DNA base pairs. We analyzed the hydrogen bond
lengths, the so-called twist angle ($\vartheta$) and the rise (R)
parameter, calculated in the same way as done in previous works\cite{gil2016theoretical}
(see Supporting Information for definition), and we compared them to the counterparts of
the original 1n37 PDB structure. Such values are depicted in Table
\ref{tbl:Duplex-geometries}.

\begin{table}
\caption{Hydrogen bond lengths, rise (R) parameter and twist angle ({\small{}$\vartheta$}) of the original 1n37 PDB structure in parentheses, after LS-DFT optimization at LMKLL/D2DZ level with SIESTA in bold, after optimization with PM6-DH2
in italics, after optimization with PM7 in italics plus bold and finally, QM/MM optimization at M11-L/6-31+G(d,p):AMBER level in normal script (see Figure S1 of the Supporting Information to check the labels of the atoms corresponding to the hydrogen bonds).}
\label{tbl:Duplex-geometries}{\small{} }%
\begin{tabular}{ccccc}
{\scriptsize{}Base Pairs  } & {\scriptsize{}Purine···Pyrimidine } & {\scriptsize{}Distance ($\textrm{Å}$)} & {\scriptsize{}R ($\textrm{Å}$)} & {\scriptsize{}\textgreek{j} (º)}\tabularnewline
\hline 
\multirow{3}{*}{{\scriptsize{}A}\textsubscript{{\scriptsize{}1}}{\scriptsize{}-T}\textsubscript{{\scriptsize{}8}}} & {\scriptsize{}N}\textsubscript{{\scriptsize{}6}}{\scriptsize{}···O}\textsubscript{{\scriptsize{}4}}{\scriptsize{}
 } & {\scriptsize{}(2.93)/}\textbf{\scriptsize{}3.16}{\scriptsize{}/}\emph{\scriptsize{}2.78}{\scriptsize{}/}\textbf{\emph{\scriptsize{}2.98}}{\scriptsize{}/2.88} &  & \tabularnewline
 & {\scriptsize{}N}\textsubscript{{\scriptsize{}1}}{\scriptsize{}···N}\textsubscript{{\scriptsize{}3}}{\scriptsize{}
 } & {\scriptsize{}(2.84)/}\textbf{\scriptsize{}2.85}{\scriptsize{}/}\emph{\scriptsize{}3.12}{\scriptsize{}/}\textbf{\emph{\scriptsize{}3.15}}{\scriptsize{}/2.92} &  & \tabularnewline
 & {\scriptsize{}C}\textsubscript{{\scriptsize{}2}}{\scriptsize{}···O}\textsubscript{{\scriptsize{}2}}{\scriptsize{}
 } & {\scriptsize{}(3.47)/}\textbf{\scriptsize{}3.38}{\scriptsize{}/}\emph{\scriptsize{}4.38}{\scriptsize{}/}\textbf{\emph{\scriptsize{}3.97}}{\scriptsize{}/3.66} &  & \tabularnewline
 &  &  & {\scriptsize{}(2.85)/}\textbf{\scriptsize{}3.61}{\scriptsize{}/}\emph{\scriptsize{}3.44}{\scriptsize{}/}\textbf{\emph{\scriptsize{}3.73}}{\scriptsize{}/3.53} & {\scriptsize{}(23.9)/}\textbf{\scriptsize{}23.2}{\scriptsize{}/}\emph{\scriptsize{}27.2}{\scriptsize{}/}\textbf{\emph{\scriptsize{}8.0}}{\scriptsize{}/22.2}\tabularnewline
\multirow{3}{*}{{\scriptsize{}G}\textsubscript{{\scriptsize{}2}}{\scriptsize{}-C}\textsubscript{{\scriptsize{}7}}} & {\scriptsize{}O}\textsubscript{{\scriptsize{}6}}{\scriptsize{}···N}\textsubscript{{\scriptsize{}4 }}{\scriptsize{} } & {\scriptsize{}(2.83)/}\textbf{\scriptsize{}2.92}{\scriptsize{}/}\emph{\scriptsize{}2.87}{\scriptsize{}/}\textbf{\emph{\scriptsize{}3.02}}{\scriptsize{}/2.89} &  & \tabularnewline
 & {\scriptsize{}N}\textsubscript{{\scriptsize{}1}}{\scriptsize{}···N}\textsubscript{{\scriptsize{}3}}{\scriptsize{} } & {\scriptsize{}(2.90)/}\textbf{\scriptsize{}2.95}{\scriptsize{}/}\emph{\scriptsize{}2.86}{\scriptsize{}/}\textbf{\emph{\scriptsize{}2.93}}{\scriptsize{}/2.90} &  & \tabularnewline
 & {\scriptsize{} N}\textsubscript{{\scriptsize{}2}}{\scriptsize{}···O}\textsubscript{{\scriptsize{}2}}{\scriptsize{}
 } & {\scriptsize{}(2.84)/}\textbf{\scriptsize{}2.93}{\scriptsize{}/}\emph{\scriptsize{}2.83}{\scriptsize{}/}\textbf{\emph{\scriptsize{}2.83}}{\scriptsize{}/2.78} &  & \tabularnewline
 &  &  & {\scriptsize{}(3.60)/}\textbf{\scriptsize{}3.15}{\scriptsize{}/}\emph{\scriptsize{}3.47}{\scriptsize{}/}\textbf{\emph{\scriptsize{}1.42}}{\scriptsize{}/2.90 } & {\scriptsize{}(23.3)/}\textbf{\scriptsize{}21.8}{\scriptsize{}/}\emph{\scriptsize{}26.2}{\scriptsize{}/}\emph{\scriptsize{}-}\textbf{\emph{\scriptsize{}43.4}}{\scriptsize{}/30.5}\tabularnewline
\multirow{3}{*}{{\scriptsize{}A}\textsubscript{{\scriptsize{}3}}{\scriptsize{}-T}\textsubscript{{\scriptsize{}6}}} & {\scriptsize{}N}\textsubscript{{\scriptsize{}6}}{\scriptsize{}···O}\textsubscript{{\scriptsize{}4}}{\scriptsize{}
 } & {\scriptsize{}(2.87)/}\textbf{\scriptsize{}3.11}{\scriptsize{}/}\emph{\scriptsize{}2.85}{\scriptsize{}/}\textbf{\emph{\scriptsize{}5.61}}{\scriptsize{}/3.02} &  & \tabularnewline
 & {\scriptsize{}N}\textsubscript{{\scriptsize{}1}}{\scriptsize{}···N}\textsubscript{{\scriptsize{}3}}{\scriptsize{}
 } & {\scriptsize{}(2.94)/}\textbf{\scriptsize{}3.00}{\scriptsize{}/}\emph{\scriptsize{}3.00}{\scriptsize{}/}\textbf{\emph{\scriptsize{}7.3}}\emph{\scriptsize{}3}{\scriptsize{}/2.87} &  & \tabularnewline
 & {\scriptsize{}C}\textsubscript{{\scriptsize{}2}}{\scriptsize{}···O}\textsubscript{{\scriptsize{}2}}{\scriptsize{}
 } & {\scriptsize{}(3.70)/}\textbf{\scriptsize{}3.84}{\scriptsize{}/}\emph{\scriptsize{}4.00}{\scriptsize{}/}\textbf{\emph{\scriptsize{}8.84}}{\scriptsize{}/3.45} &  & \tabularnewline
 &  &  & {\scriptsize{}(3.54)/}\textbf{\scriptsize{}3.46}{\scriptsize{}/}\emph{\scriptsize{}3.54}{\scriptsize{}/}\textbf{\emph{\scriptsize{}2.89}}{\scriptsize{}/3.57 } & {\scriptsize{}(21.6)/}\textbf{\scriptsize{}21.6}{\scriptsize{}/}\emph{\scriptsize{}20.5}{\scriptsize{}/}\textbf{\emph{\scriptsize{}15.3}}{\scriptsize{}/20.9}\tabularnewline
\multirow{3}{*}{{\scriptsize{}C}\textsubscript{{\scriptsize{}4}}{\scriptsize{}-G}\textsubscript{{\scriptsize{}5}}{\scriptsize{} }} & {\scriptsize{}O}\textsubscript{{\scriptsize{}6}}{\scriptsize{}···N}\textsubscript{{\scriptsize{}4 }}{\scriptsize{} } & {\scriptsize{}(2.92)/}\textbf{\scriptsize{}3.80}{\scriptsize{}/}\emph{\scriptsize{}2.87}{\scriptsize{}/}\textbf{\emph{\scriptsize{}2.89}}{\scriptsize{}/2.95} &  & \tabularnewline
 & {\scriptsize{}N}\textsubscript{{\scriptsize{}1}}{\scriptsize{}···N}\textsubscript{{\scriptsize{}3}}{\scriptsize{} } & {\scriptsize{}(2.91)/}\textbf{\scriptsize{}3.29}{\scriptsize{}/}\emph{\scriptsize{}2.84}{\scriptsize{}/}\textbf{\emph{\scriptsize{}3.00}}{\scriptsize{}/2.91} &  & \tabularnewline
 & {\scriptsize{} N}\textsubscript{{\scriptsize{}2}}{\scriptsize{}···O}\textsubscript{{\scriptsize{}2}}{\scriptsize{}
 } & {\scriptsize{}(2.80)/}\textbf{\scriptsize{}2.87}{\scriptsize{}/}\emph{\scriptsize{}2.82}{\scriptsize{}/}\textbf{\emph{\scriptsize{}2.92}}{\scriptsize{}/2.78} &  & \tabularnewline
 &  &  & {\scriptsize{}(5.66)/}\textbf{\scriptsize{}5.75}{\scriptsize{}/}\emph{\scriptsize{}5.94}{\scriptsize{}/}\textbf{\emph{\scriptsize{}5.51}}{\scriptsize{}/6.05 } & {\scriptsize{}(5.2)/}\textbf{\scriptsize{}24.5}{\scriptsize{}/}\emph{\scriptsize{}30.5}{\scriptsize{}/}\textbf{\emph{\scriptsize{}15.4}}{\scriptsize{}/21.9}\tabularnewline
\multirow{3}{*}{{\scriptsize{}G}\textsubscript{{\scriptsize{}5}}{\scriptsize{}-C}\textsubscript{{\scriptsize{}4}}} & {\scriptsize{}O}\textsubscript{{\scriptsize{}6}}{\scriptsize{}···N}\textsubscript{{\scriptsize{}4 }}{\scriptsize{} } & {\scriptsize{}(2.87)/}\textbf{\scriptsize{}2.93}{\scriptsize{}/}\emph{\scriptsize{}2.85}{\scriptsize{}/}\textbf{\emph{\scriptsize{}2.88}}{\scriptsize{}/2.84} &  & \tabularnewline
 & {\scriptsize{}N}\textsubscript{{\scriptsize{}1}}{\scriptsize{}···N}\textsubscript{{\scriptsize{}3}}{\scriptsize{} } & {\scriptsize{} (2.88)/2}\textbf{\scriptsize{}.99}{\scriptsize{}/}\emph{\scriptsize{}2.87}{\scriptsize{}/}\textbf{\emph{\scriptsize{}2.91}}{\scriptsize{}/2.90} &  & \tabularnewline
 & {\scriptsize{} N}\textsubscript{{\scriptsize{}2}}{\scriptsize{}···O}\textsubscript{{\scriptsize{}2}}{\scriptsize{}
 } & {\scriptsize{}(2.83)/}\textbf{\scriptsize{}2.96}{\scriptsize{}/}\emph{\scriptsize{}2.83}{\scriptsize{}/}\textbf{\emph{\scriptsize{}2,86}}{\scriptsize{}/2.84} &  & \tabularnewline
 &  &  & {\scriptsize{}(3.52)/}\textbf{\scriptsize{}3.31}{\scriptsize{}/}\emph{\scriptsize{}3.16}{\scriptsize{}/}\textbf{\emph{\scriptsize{}3.10}}{\scriptsize{}/3.35 } & {\scriptsize{}(25.0)/}\textbf{\scriptsize{}22.7}{\scriptsize{}/}\emph{\scriptsize{}26.2}{\scriptsize{}/}\textbf{\emph{\scriptsize{}28.4}}{\scriptsize{}/23.6}\tabularnewline
\multirow{3}{*}{{\scriptsize{}T}\textsubscript{{\scriptsize{}6}}{\scriptsize{}-A}\textsubscript{{\scriptsize{}3}}} & {\scriptsize{}N}\textsubscript{{\scriptsize{}6}}{\scriptsize{}···O}\textsubscript{{\scriptsize{}4}}{\scriptsize{}
 } & {\scriptsize{}(2.96)/}\textbf{\scriptsize{}3.27}{\scriptsize{}/}\emph{\scriptsize{}2.84}{\scriptsize{}/}\textbf{\emph{\scriptsize{}3.25}}{\scriptsize{}/2.89} &  & \tabularnewline
 & {\scriptsize{}N}\textsubscript{{\scriptsize{}1}}{\scriptsize{}···N}\textsubscript{{\scriptsize{}3}}{\scriptsize{}
 } & {\scriptsize{}(2.89)/}\textbf{\scriptsize{}3.04}{\scriptsize{}/}\emph{\scriptsize{}3.05}{\scriptsize{}/}\textbf{\emph{\scriptsize{}2.89}}{\scriptsize{}/2.86} &  & \tabularnewline
 & {\scriptsize{}C}\textsubscript{{\scriptsize{}2}}{\scriptsize{}···O}\textsubscript{{\scriptsize{}2}}{\scriptsize{}
 } & {\scriptsize{}(3.61)/}\textbf{\scriptsize{}3.79}{\scriptsize{}/}\emph{\scriptsize{}4.18}{\scriptsize{}/}\textbf{\emph{\scriptsize{}3.35}}{\scriptsize{}/3.46} &  & \tabularnewline
 &  &  & {\scriptsize{}(3.15)/}\textbf{\scriptsize{}3.07}{\scriptsize{}/}\emph{\scriptsize{}3.31}{\scriptsize{}/}\textbf{\emph{\scriptsize{}2.90}}{\scriptsize{}/4.02 } & {\scriptsize{}(26.8)/}\textbf{\scriptsize{}26.5}{\scriptsize{}/}\emph{\scriptsize{}26.2}{\scriptsize{}/}\textbf{\emph{\scriptsize{}20.2}}{\scriptsize{}/29.7}\tabularnewline
\multirow{3}{*}{{\scriptsize{}C}\textsubscript{{\scriptsize{}7}}{\scriptsize{}-G}\textsubscript{{\scriptsize{}2}}} & {\scriptsize{}O}\textsubscript{{\scriptsize{}6}}{\scriptsize{}···N}\textsubscript{{\scriptsize{}4 }}{\scriptsize{} } & {\scriptsize{}(2.80)/}\textbf{\scriptsize{}2.88}{\scriptsize{}/}\emph{\scriptsize{}2.83}{\scriptsize{}/}\textbf{\emph{\scriptsize{}2.94}}{\scriptsize{}/2.87} &  & \tabularnewline
 & {\scriptsize{}N}\textsubscript{{\scriptsize{}1}}{\scriptsize{}···N}\textsubscript{{\scriptsize{}3}}{\scriptsize{} } & {\scriptsize{}(2.94)/}\textbf{\scriptsize{}3.01}{\scriptsize{}/}\emph{\scriptsize{}2.85}{\scriptsize{}/}\textbf{\emph{\scriptsize{}2.98}}{\scriptsize{}/2.85} &  & \tabularnewline
 & {\scriptsize{} N}\textsubscript{{\scriptsize{}2}}{\scriptsize{}···O}\textsubscript{{\scriptsize{}2}}{\scriptsize{}
 } & {\scriptsize{} (2.99)/}\textbf{\scriptsize{}3.08}{\scriptsize{}/}\emph{\scriptsize{}2.86}{\scriptsize{}/}\textbf{\emph{\scriptsize{}2.97}}{\scriptsize{}/2.89} &  & \tabularnewline
 &  &  & {\scriptsize{}(3.18)/}\textbf{\scriptsize{}3.14}{\scriptsize{}/}\emph{\scriptsize{}2.73}{\scriptsize{}/}\textbf{\emph{\scriptsize{}0.44}}{\scriptsize{}/2.79 } & {\scriptsize{}(0.7)/}\textbf{\scriptsize{}-0.9}{\scriptsize{}/}\emph{\scriptsize{}-12.5}{\scriptsize{}/}\textbf{\emph{\scriptsize{}18.8}}{\scriptsize{}/-17.0}\tabularnewline
\multirow{3}{*}{{\scriptsize{}T}\textsubscript{{\scriptsize{}8}}{\scriptsize{}-A}\textsubscript{{\scriptsize{}1}}} & {\scriptsize{}N}\textsubscript{{\scriptsize{}6}}{\scriptsize{}···O}\textsubscript{{\scriptsize{}4}}{\scriptsize{}
 } & {\scriptsize{}(7.40)/}\textbf{\scriptsize{}7.86}{\scriptsize{}/}\emph{\scriptsize{}3.04}{\scriptsize{}/}\textbf{\emph{\scriptsize{}5.41}}{\scriptsize{}/3.46} &  & \tabularnewline
 & {\scriptsize{}N}\textsubscript{{\scriptsize{}1}}{\scriptsize{}···N}\textsubscript{{\scriptsize{}3}}{\scriptsize{}
 } & {\scriptsize{}(7.31)/}\textbf{\scriptsize{}7.89}{\scriptsize{}/}\emph{\scriptsize{}4.67}{\scriptsize{}/}\textbf{\emph{\scriptsize{}6.08}}{\scriptsize{}/7.93} &  & \tabularnewline
 & {\scriptsize{}C}\textsubscript{{\scriptsize{}2}}{\scriptsize{}···O}\textsubscript{{\scriptsize{}2}}{\scriptsize{}
 } & {\scriptsize{}(6.12)/}\textbf{\scriptsize{}6.68}{\scriptsize{}/}\emph{\scriptsize{}7.37}{\scriptsize{}/}\textbf{\emph{\scriptsize{}7.57}}{\scriptsize{}/11.19} &  & \tabularnewline
\hline 
\end{tabular}
\end{table}

First, it is worth mentioning that the PM7 semi-empirical method gives the worst geometrical results for the 1n37 system. Although the main structural features are kept, some of the base pairs, such as A\textsubscript{3}-T\textsubscript{6} or
T\textsubscript{8}-A\textsubscript{1}, are separated. The structure is not kept straight, in a conformation with all bases stacked, but curved in a \textquotedbl C\textquotedbl{} shape. For these reasons, we will not further consider this structure for deeper geometrical analysis.

Let us start analyzing the twist angle parameter, calculated in the same way as in previous works.\cite{galliot2017effects} LMKLL is the method that gives the best results for the twist angle, with differences from 0.0º to 1.3º in most cases. Only two twist angles differ by more than 2º: the twist angle formed between C\textsubscript{4}-G\textsubscript{5} and G\textsubscript{5}-C\textsubscript{4} and that formed between G\textsubscript{5}-C\textsubscript{4} and T\textsubscript{6}-A\textsubscript{3}. On the other hand, it is unclear which approach gives the worst results for the twist angle.  In some cases, QM/MM yields the worst results, whereas in others, PM6-DH2 does. In general, the results for the twist angle given by the three different approaches agree with the analysis of the RMSD.

Regarding the Rise parameter (R), the QM/MM approach at M11-L/6-31+G(d,p):
AMBER level of theory gives the worst results with differences to the PDB structure higher than 0.5 $\textrm{Å}$ in some cases. On the other hand, as a general trend, the best results are again given by the LS-DFT. Nevertheless, the SE approach also gives excellent results for the rise parameter (R), in some cases (A\textsubscript{1}-T\textsubscript{8}/G\textsubscript{2}-C\textsubscript{7},
G\textsubscript{2}-C\textsubscript{7}/A\textsubscript{3}-T\textsubscript{6},
and A\textsubscript{3}-T\textsubscript{6}/C\textsubscript{4}-G\textsubscript{5}), even better than the LS-DFT approach. Thus, it is not clear which of the two approaches, LS-DFT or semi-empirical, including dispersion yield the best results. 

Interesting trends are also observed for hydrogen bond distances. In the case of the hydrogen bond distances of the A-T base pairs, we see that the SE approach gives the worst behavior. Indeed, there is a considerable lengthening, especially for the N\textsubscript{1}···H-N\textsubscript{3} and C\textsubscript{2}-H···O\textsubscript{2} hydrogen bonds. LS-DFT approach gives the best results for the hydrogen bond distances again, although in general, they are elongated with respect to the PDB structure,  especially for the N\textsubscript{1}···H-N\textsubscript{3} and C\textsubscript{2}-H···O\textsubscript{2} hydrogen bonds. Nevertheless, it must be mentioned that the QM/MM approach at M11-L/6-31+G(d,p):AMBER
also gives the best results in some of the N\textsubscript{6}-H···O\textsubscript{4}
hydrogen bonds.  On the other hand, there is a better agreement between theoretical hydrogen bond distances and those corresponding to the original PDB structure for the  G-C base pairs, with LS-DFT giving the best results.  

Summarizing, the semi-empirical approach with the PM6-DH2 Hamiltonian gives in some cases better results than the most popular QM/MM approach at M11-L/6-31+G(d,p):AMBER level of theory. On the other hand, the LS-DFT method gives an excellent agreement between the geometrical parameters of the PDB structure and those of the optimized structure. SIESTA software gives us excellent results for this biomolecular system of 665 atoms in a reasonable computation time.

\subsubsection{G-quadruplex}

\begin{figure}
\includegraphics[scale=0.65]{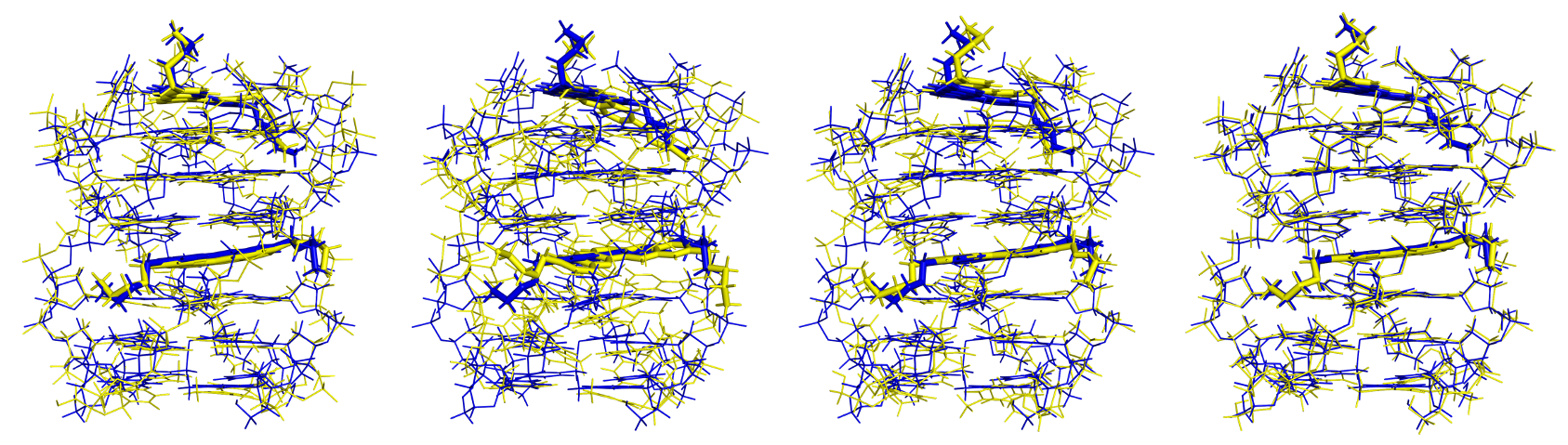}

\caption{Overlapping of 2jwq PDB structure (in blue) with the optimized geometries (in yellow) with the PM6-DH2, PM7, B3LYP-D3(GD3BJ)/6-31+G(d,p):AMBER and LMKLL methods, from left to right. RMSD values of 0.98 $\textrm{Å}$, 2.06 $\textrm{Å}$, 1.05 $\textrm{Å}$ and 0.24 $\textrm{Å}$, respectively.}

\label{fgr:G-quadruplex_structures} 
\end{figure}

We carried out an analysis for the systems based on GQ by using similar structural parameters as for the dDNA (See Table \ref{tbl:G-quadruplex_geometries}): RMSD from the PDB reference structure, hydrogen bond lengths, rise distances, and twist angle, in this case as defined by Chung et al.\cite{chung2015structure}(see Supporting Information for the nomenclature in Figure S3, Rise definition, and twist angle definition in Figure S2). The 2jwq PDB structure for the GQ was optimized at PM6-DH2 and PM7 level (in the case of the semi-empirical approach), at B3LYP-D3(GD3BJ)/6-31+G(d,p):AMBER level (in the case of the QM/MM approach) and at LMKLL/DZDP level (an  LS-DFT method), see Figure \ref{fgr:G-quadruplex_structures}. Semi-empirical methods yielded higher RMSD values than the LS-DFT method. The PM6-DH2 and PM7 methods gave RMSD values of 0.98 and 2.06 Å, respectively. Quite interestingly, the RMSD values obtained with the semi-empirical models for the systems based on GQ were lower than those obtained for dDNA, even though the GQ used in this work have approximately twice as many atoms as the studied dDNA systems. 
This better agreement could be due to a more rigid structure of the GQ because the stacking of the G-tetrads creates a more rigid structure, which in turn increase the rigidity of the side loops.

\begin{table}
\caption{Hydrogen bond lengths, rise (R) parameter and twist angle ($\vartheta$) of the original 2jwq PDB structure in parentheses, after LS-DFT optimization at LMKLL/D2DZ level with SIESTA in bold, after optimization with PM6-DH2 in italics, after optimization with PM7 semi-empirical method in italic plus bold, and at the B3LYP-D3(GD3BJ)/6-31+G(d,p):AMBER level in normal script (see Figure S3 of the Supporting Information to check the labels of the atoms corresponding to the hydrogen bonds and the labels for guanines in the tetrads).}
\label{tbl:G-quadruplex_geometries}{\small{}}%
\begin{tabular}{ccccc}
{\scriptsize{}Tetrads  } & {\scriptsize{}Guanine···Guanine } & {\scriptsize{}Distance ($\textrm{Å}$)} & {\scriptsize{}R ($\textrm{Å}$)} & {\scriptsize{}\textgreek{j} (º)}\tabularnewline
\hline 
\multirow{8}{*}{{\scriptsize{}G}\textsubscript{{\scriptsize{}4,11,18,25}}} & {\scriptsize{}G}\textsubscript{{\scriptsize{}4}}{\scriptsize{}O}\textsubscript{{\scriptsize{}6}}{\scriptsize{}···G}\textsubscript{{\scriptsize{}18}}{\scriptsize{}N}\textsubscript{{\scriptsize{}1}}{\scriptsize{}
 } & {\scriptsize{}(3.01)/}\textbf{\scriptsize{}2.90}{\scriptsize{}/}\emph{\scriptsize{}2.90}{\scriptsize{}/}\textbf{\emph{\scriptsize{}3.00}}{\scriptsize{}/2.92} &  & \tabularnewline
 & {\scriptsize{}G}\textsubscript{{\scriptsize{}4}}{\scriptsize{}N}\textsubscript{{\scriptsize{}7}}{\scriptsize{}···G}\textsubscript{{\scriptsize{}18}}{\scriptsize{}N}\textsubscript{{\scriptsize{}2}}{\scriptsize{} } & {\scriptsize{}(2.88)/}\textbf{\scriptsize{}3.00}{\scriptsize{}/}\emph{\scriptsize{}3.04}{\scriptsize{}/}\textbf{\emph{\scriptsize{}3.01}}{\scriptsize{}/2.95} &  & \tabularnewline
 & {\scriptsize{}G}\textsubscript{{\scriptsize{}18}}{\scriptsize{}O}\textsubscript{{\scriptsize{}6}}{\scriptsize{}···G}\textsubscript{{\scriptsize{}11}}{\scriptsize{}N}\textsubscript{{\scriptsize{}1}}{\scriptsize{} } & {\scriptsize{}(3.14)/}\textbf{\scriptsize{}2.94}{\scriptsize{}/}\emph{\scriptsize{}2.88}{\scriptsize{}/}\textbf{\emph{\scriptsize{}2.96}}{\scriptsize{}/2.94} &  & \tabularnewline
 & {\scriptsize{} G}\textsubscript{{\scriptsize{}18}}{\scriptsize{}N}\textsubscript{{\scriptsize{}7}}{\scriptsize{}···G}\textsubscript{{\scriptsize{}11}}{\scriptsize{}N}\textsubscript{{\scriptsize{}2}}{\scriptsize{} } & {\scriptsize{}(2.77)/}\textbf{\scriptsize{}2.93}{\scriptsize{}/}\emph{\scriptsize{}3.00}{\scriptsize{}/}\textbf{\emph{\scriptsize{}3.07}}{\scriptsize{}/2.88} &  & \tabularnewline
 & {\scriptsize{}G}\textsubscript{{\scriptsize{}11}}{\scriptsize{}O}\textsubscript{{\scriptsize{}6}}{\scriptsize{}···G}\textsubscript{{\scriptsize{}25}}{\scriptsize{}N}\textsubscript{{\scriptsize{}1}}{\scriptsize{} } & {\scriptsize{}(2.85)/}\textbf{\scriptsize{}2.90}{\scriptsize{}/}\emph{\scriptsize{}2.90}{\scriptsize{}/}\textbf{\emph{\scriptsize{}2.81}}{\scriptsize{}/2.89} &  & \tabularnewline
 & {\scriptsize{}G}\textsubscript{{\scriptsize{}11}}{\scriptsize{}N}\textsubscript{{\scriptsize{}7}}{\scriptsize{}···G}\textsubscript{{\scriptsize{}25}}{\scriptsize{}N}\textsubscript{{\scriptsize{}2}}{\scriptsize{} } & {\scriptsize{}(3.41)/}\textbf{\scriptsize{}3.16}{\scriptsize{}/}\emph{\scriptsize{}3.06}{\scriptsize{}/}\textbf{\emph{\scriptsize{}3.15}}{\scriptsize{}/3.01} &  & \tabularnewline
 & {\scriptsize{}G}\textsubscript{{\scriptsize{}25}}{\scriptsize{}O}\textsubscript{{\scriptsize{}6}}{\scriptsize{}···G}\textsubscript{{\scriptsize{}4}}{\scriptsize{}N}\textsubscript{{\scriptsize{}1}}{\scriptsize{} } & {\scriptsize{}(2.92)/}\textbf{\scriptsize{}2.91}{\scriptsize{}/}\emph{\scriptsize{}2.82}{\scriptsize{}/}\textbf{\emph{\scriptsize{}2.88}}{\scriptsize{}/2.89} &  & \tabularnewline
 & {\scriptsize{}G}\textsubscript{{\scriptsize{}25}}{\scriptsize{}N}\textsubscript{{\scriptsize{}7}}{\scriptsize{}···G}\textsubscript{{\scriptsize{}4}}{\scriptsize{}N}\textsubscript{{\scriptsize{}2}}{\scriptsize{} } & {\scriptsize{}(2.93)/}\textbf{\scriptsize{}3.02}{\scriptsize{}/}\emph{\scriptsize{}2.97}{\scriptsize{}/}\textbf{\emph{\scriptsize{}3.06}}{\scriptsize{}/2.88} &  & \tabularnewline
 &  &  & {\scriptsize{}(3.38)/}\textbf{\scriptsize{}3.41}{\scriptsize{}/}\emph{\scriptsize{}3.31}{\scriptsize{}/}\textbf{\emph{\scriptsize{}3.33}}{\scriptsize{}/3.39} & {\scriptsize{}(28.7)/}\textbf{\scriptsize{}25.3}{\scriptsize{}/}\emph{\scriptsize{}26.8}{\scriptsize{}/}\textbf{\emph{\scriptsize{}25.3}}{\scriptsize{}/24.8}\tabularnewline
\multirow{8}{*}{{\scriptsize{}G}\textsubscript{{\scriptsize{}5,12,19,26}}} & {\scriptsize{}G}\textsubscript{{\scriptsize{}5}}{\scriptsize{}O}\textsubscript{{\scriptsize{}6}}{\scriptsize{}···G}\textsubscript{{\scriptsize{}19}}{\scriptsize{}N}\textsubscript{{\scriptsize{}1}}{\scriptsize{}
 } & {\scriptsize{}(3.10)/}\textbf{\scriptsize{}2.91}{\scriptsize{}/}\emph{\scriptsize{}2.74}{\scriptsize{}/}\textbf{\emph{\scriptsize{}2.82}}{\scriptsize{}/3.17} &  & \tabularnewline
 & {\scriptsize{}G}\textsubscript{{\scriptsize{}5}}{\scriptsize{}N}\textsubscript{{\scriptsize{}7}}{\scriptsize{}···G}\textsubscript{{\scriptsize{}19}}{\scriptsize{}N}\textsubscript{{\scriptsize{}2}}{\scriptsize{} } & {\scriptsize{}(3.45)/}\textbf{\scriptsize{}2.98}{\scriptsize{}/}\emph{\scriptsize{}3.20}{\scriptsize{}/}\textbf{\emph{\scriptsize{}2.96}}{\scriptsize{}/2.92} &  & \tabularnewline
 & {\scriptsize{}G}\textsubscript{{\scriptsize{}19}}{\scriptsize{}O}\textsubscript{{\scriptsize{}6}}{\scriptsize{}···G}\textsubscript{{\scriptsize{}12}}{\scriptsize{}N}\textsubscript{{\scriptsize{}1}}{\scriptsize{} } & {\scriptsize{}(2.53)/}\textbf{\scriptsize{}3.08}{\scriptsize{}/}\emph{\scriptsize{}2.86}{\scriptsize{}/}\textbf{\emph{\scriptsize{}2.87}}{\scriptsize{}/2.98} &  & \tabularnewline
 & {\scriptsize{}G}\textsubscript{{\scriptsize{}19}}{\scriptsize{}N}\textsubscript{{\scriptsize{}7}}{\scriptsize{}···G}\textsubscript{{\scriptsize{}12}}{\scriptsize{}N}\textsubscript{{\scriptsize{}2}}{\scriptsize{} } & {\scriptsize{}(2.43)/}\textbf{\scriptsize{}2.98}{\scriptsize{}/}\emph{\scriptsize{}3.14}{\scriptsize{}/}\textbf{\emph{\scriptsize{}3.01}}{\scriptsize{}/2.86} &  & \tabularnewline
 & {\scriptsize{}G}\textsubscript{{\scriptsize{}12}}{\scriptsize{}O}\textsubscript{{\scriptsize{}6}}{\scriptsize{}···G}\textsubscript{{\scriptsize{}26}}{\scriptsize{}N}\textsubscript{{\scriptsize{}1}}{\scriptsize{} } & {\scriptsize{}(2.91)/}\textbf{\scriptsize{}2.85}{\scriptsize{}/}\emph{\scriptsize{}2.79}{\scriptsize{}/}\textbf{\emph{\scriptsize{}2.98}}{\scriptsize{}/2.95} &  & \tabularnewline
 & {\scriptsize{}G}\textsubscript{{\scriptsize{}12}}{\scriptsize{}N}\textsubscript{{\scriptsize{}7}}{\scriptsize{}···G}\textsubscript{{\scriptsize{}26}}{\scriptsize{}N}\textsubscript{{\scriptsize{}2}}{\scriptsize{} } & {\scriptsize{}(3.77)/}\textbf{\scriptsize{}3.38}{\scriptsize{}/}\emph{\scriptsize{}3.53}{\scriptsize{}/}\textbf{\emph{\scriptsize{}2.96}}{\scriptsize{}/3.13} &  & \tabularnewline
 & {\scriptsize{}G}\textsubscript{{\scriptsize{}26}}{\scriptsize{}O}\textsubscript{{\scriptsize{}6}}{\scriptsize{}···G}\textsubscript{{\scriptsize{}5}}{\scriptsize{}N}\textsubscript{{\scriptsize{}1}}{\scriptsize{} } & {\scriptsize{}(3.08)/}\textbf{\scriptsize{}2.92}{\scriptsize{}/}\emph{\scriptsize{}2.80}{\scriptsize{}/}\textbf{\emph{\scriptsize{}2.89}}{\scriptsize{}/2.90} &  & \tabularnewline
 & {\scriptsize{}G}\textsubscript{{\scriptsize{}26}}{\scriptsize{}N}\textsubscript{{\scriptsize{}7}}{\scriptsize{}···G}\textsubscript{{\scriptsize{}5}}{\scriptsize{}N}\textsubscript{{\scriptsize{}2}}{\scriptsize{} } & {\scriptsize{}(3.17)/}\textbf{\scriptsize{}3.00}{\scriptsize{}/}\emph{\scriptsize{}3.06}{\scriptsize{}/}\textbf{\emph{\scriptsize{}2.89}}{\scriptsize{}/2.96} &  & \tabularnewline
 &  &  & {\scriptsize{}(3.19)/}\textbf{\scriptsize{}3.26}{\scriptsize{}/}\emph{\scriptsize{}3.17}{\scriptsize{}/}\textbf{\emph{\scriptsize{}3.21}}{\scriptsize{}/3.42} & {\scriptsize{}(16.4)/}\textbf{\scriptsize{}21.9}{\scriptsize{}/}\emph{\scriptsize{}19.5}{\scriptsize{}/}\textbf{\emph{\scriptsize{}30.3}}{\scriptsize{}/28.5}\tabularnewline
\multirow{8}{*}{{\scriptsize{}G}\textsubscript{{\scriptsize{}6,13,20,27}}{\scriptsize{} }} & {\scriptsize{}G}\textsubscript{{\scriptsize{}6}}{\scriptsize{}O}\textsubscript{{\scriptsize{}6}}{\scriptsize{}···G}\textsubscript{{\scriptsize{}20}}{\scriptsize{}N}\textsubscript{{\scriptsize{}1}}{\scriptsize{}
 } & {\scriptsize{}(2.83)/}\textbf{\scriptsize{}2.85}{\scriptsize{}/}\emph{\scriptsize{}2.77}{\scriptsize{}/}\textbf{\emph{\scriptsize{}2.88}}{\scriptsize{}/2.75} &  & \tabularnewline
 & {\scriptsize{}G}\textsubscript{{\scriptsize{}6}}{\scriptsize{}N}\textsubscript{{\scriptsize{}7}}{\scriptsize{}···G}\textsubscript{{\scriptsize{}20}}{\scriptsize{}N}\textsubscript{{\scriptsize{}2}}{\scriptsize{} } & {\scriptsize{}(2.85)/}\textbf{\scriptsize{}3.02}{\scriptsize{}/}\emph{\scriptsize{}3.06}{\scriptsize{}/}\textbf{\emph{\scriptsize{}2.93}}{\scriptsize{}/3.02} &  & \tabularnewline
 & {\scriptsize{}G}\textsubscript{{\scriptsize{}20}}{\scriptsize{}O}\textsubscript{{\scriptsize{}6}}{\scriptsize{}···G}\textsubscript{{\scriptsize{}13}}{\scriptsize{}N}\textsubscript{{\scriptsize{}1}}{\scriptsize{} } & {\scriptsize{}(2.99)/}\textbf{\scriptsize{}2.99}{\scriptsize{}/}\emph{\scriptsize{}2.80}{\scriptsize{}/}\textbf{\emph{\scriptsize{}2.87}}{\scriptsize{}/2.82} &  & \tabularnewline
 & {\scriptsize{}G}\textsubscript{{\scriptsize{}20}}{\scriptsize{}N}\textsubscript{{\scriptsize{}7}}{\scriptsize{}···G}\textsubscript{{\scriptsize{}13}}{\scriptsize{}N}\textsubscript{{\scriptsize{}2}}{\scriptsize{} } & {\scriptsize{}(2.80)/}\textbf{\scriptsize{}3.00}{\scriptsize{}/}\emph{\scriptsize{}3.18}{\scriptsize{}/}\textbf{\emph{\scriptsize{}3.05}}{\scriptsize{}/2.89} &  & \tabularnewline
 & {\scriptsize{}G}\textsubscript{{\scriptsize{}13}}{\scriptsize{}O}\textsubscript{{\scriptsize{}6}}{\scriptsize{}···G}\textsubscript{{\scriptsize{}27}}{\scriptsize{}N}\textsubscript{{\scriptsize{}1}}{\scriptsize{} } & {\scriptsize{}(2.89)/}\textbf{\scriptsize{}2.99}{\scriptsize{}/}\emph{\scriptsize{}2.83}{\scriptsize{}/}\textbf{\emph{\scriptsize{}2.87}}{\scriptsize{}/2.87} &  & \tabularnewline
 & {\scriptsize{}G}\textsubscript{{\scriptsize{}13}}{\scriptsize{}N}\textsubscript{{\scriptsize{}7}}{\scriptsize{}···G}\textsubscript{{\scriptsize{}27}}{\scriptsize{}N}\textsubscript{{\scriptsize{}2}}{\scriptsize{} } & {\scriptsize{}(3.05)/}\textbf{\scriptsize{}3.04}{\scriptsize{}/}\emph{\scriptsize{}3.17}{\scriptsize{}/}\textbf{\emph{\scriptsize{}2.98}}{\scriptsize{}/2.84} &  & \tabularnewline
 & {\scriptsize{}G}\textsubscript{{\scriptsize{}27}}{\scriptsize{}O}\textsubscript{{\scriptsize{}6}}{\scriptsize{}···G}\textsubscript{{\scriptsize{}6}}{\scriptsize{}N}\textsubscript{{\scriptsize{}1}}{\scriptsize{} } & {\scriptsize{}(2.86)/}\textbf{\scriptsize{}2.86}{\scriptsize{}/}\emph{\scriptsize{}2.80}{\scriptsize{}/}\textbf{\emph{\scriptsize{}2.93}}{\scriptsize{}/2.85} &  & \tabularnewline
 & {\scriptsize{}G}\textsubscript{{\scriptsize{}27}}{\scriptsize{}N}\textsubscript{{\scriptsize{}7}}{\scriptsize{}···G}\textsubscript{{\scriptsize{}6}}{\scriptsize{}N}\textsubscript{{\scriptsize{}2}}{\scriptsize{} } & {\scriptsize{}(2.91)/}\textbf{\scriptsize{}3.04}{\scriptsize{}/}\emph{\scriptsize{}3.02}{\scriptsize{}/}\textbf{\emph{\scriptsize{}2.98}}{\scriptsize{}/2.86} &  & \tabularnewline
\hline 
\end{tabular}
\end{table}

The LMKLL linear-scaling DFT method gave the best agreement to the experimental structure again, with a RMSD of 0.24 Å.  On the other hand, QM/MM and PM6-DH2 show a similar RMSD, 1.05  Å for the former and 0.98 for the latter, whereas PM7 gives the worst result, namely, a RMSD of 2.06 Å. There is a tendency for the different computational methods to reduce the range in distances between the heavy atoms in the hydrogen bonds with respect to the values observed in the PDB. The PDB structure shows a range in \(O_{6}\)···\(N_{1}\) distances between 2.53-3.14 Å, whereas for LMKLL, QM/MM, PM6-DH2, and PM7, the ranges are 2.85-3.08 Å, 2.75-3.17 Å, 2.74-2.9 Å, and 2.81-3.00 Å, respectively. The differences are even larger in the case of the N\textsubscript{7}···N\textsubscript{2} distances, in which the PDB range is 2.43-3.77 Å while for the LMKLL, QM/MM, PM6-DH2, and PM7 are 2.93-3.38 Å, 2.82-3.13 Å, 2.97-3.53 Å and 2.89-3.15 Å, respectively. The LMKLL is again the method that better reproduces these distances, followed by QM/MM, PM6-DH2, and PM7 methods. These changes in distance, although small, cause the guanine bases to rotate directly affecting the twist angle. To calculate the twist angle we used a method devised by Phan et al.\cite{chung2015structure} The rotation angle between the G-tetrads was established as the average value obtained by calculating the angle of two guanines stacked between the vectors formed for each guanine by the coordinate of the C\textsubscript{8} atom and the midpoint of the coordinates of the N\textsubscript{1} and C\textsubscript{2} atoms. We have found that the twist angle for
G\textsubscript{4,11,18,25} - G\textsubscript{5,12,19,26} is well reproduced by all methods. However the twist angle for G\textsubscript{5,12,19,26} - G\textsubscript{6,13,20,27} (16.4º) is reproduced appropriately by LMKLL (21.9º) and QM/MM (19.5º), whereas PM6-DH2 (30.3º)  and PM7 (28.5º) semi-empirical methods show larger discrepancies.  
Finally, the Rise distance is well reproduced by all methods with differences with respect to the PDB reference structure within 0.2 \AA in all cases. 

In summary, the LS-DFT at the LMKLL/DZDP level of theory is the method that best describes the geometrical structure of the GQ system. In general, we observe a better agreement between the optimized and PDB structures for GQ than for dDNA, which can be related to the more rigid structure of the former. The geometrical results obtained for 1n37 dDNA and 2jwq GQ confirm that our LS-DFT approach is appropriate to optimize DNA-type biomolecules. On the other hand, PM6-DH2 describes reasonably well both dDNA and GQ DNA systems but not the PM7 method. In addition, the PM6-DH2 approach may yield geometries of similar quality as for the popular QM/MM methodology. However, the PM7 Hamiltonian shows the most significant deviations in geometries. Finally, it must be said that the popular QM/MM approach we used here at B3LYP-D3(GD3BJ)/6-31+G(d,p):AMBER level may also lead to qualitatively correct structures.

\subsection{Energetics discussion}

\subsubsection{DNA base pairs}
\begin{figure}
\includegraphics[scale=0.6]{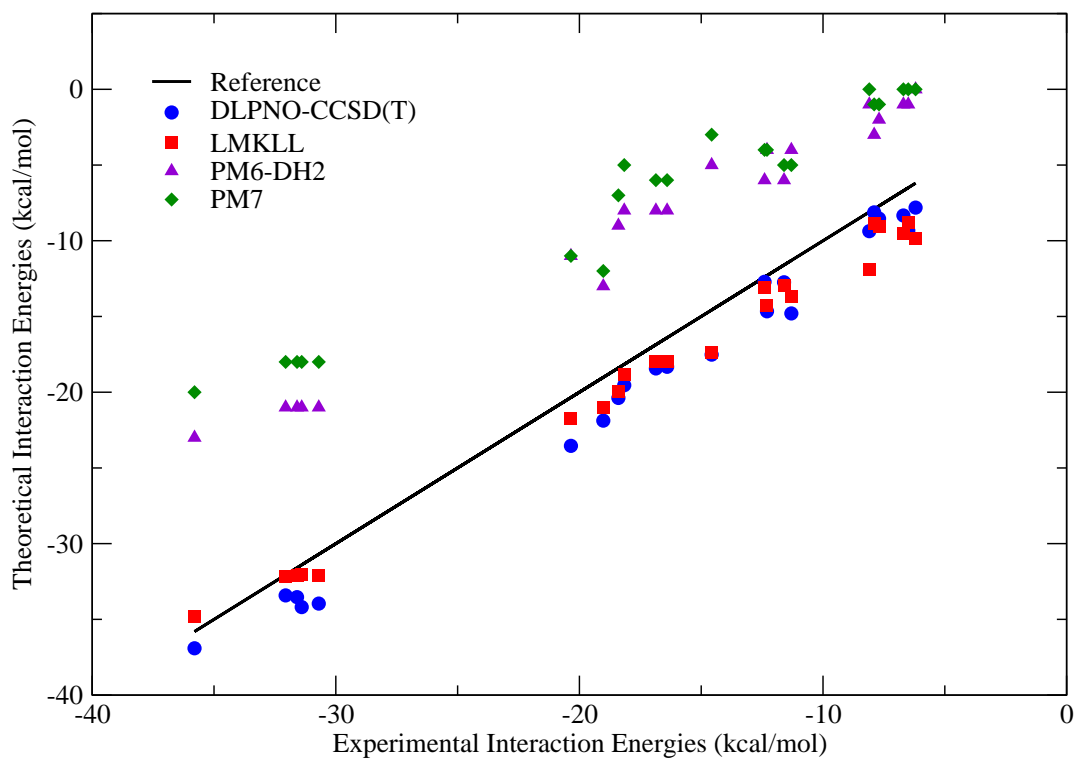}

\caption{Interaction energies (kcal/mol) for DNA base pair systems, with CCSD energies from reference 38 as the reference data. The correlation coefficients \(r^2\) for the DLPNO-CCSD(T), LMKLL, PM6-DH2, and PM7 are 0.996, 0.996, 0.990, 0.984, respectively.}

\label{fgr:energy_Pavel} 
\end{figure}

 Fig. \ref{fgr:energy_Pavel} shows graphically the trends for the interaction energies ($\Delta$E\textsubscript{int}) of different DNA hydrogen-bonded and stacked base-pairs in the gas phase for the different computational methods studied in our work. The corresponding values are depicted in Table S1. 

It is observed that the DLPNO-CCSD(T) and LMKLL methods reproduced the interaction energies of the benchmark references\textsuperscript{\cite{jurevcka2006benchmark}} accurately. 
The obtained $\Delta$E\textsubscript{int} for LMKLL and DLPNO-CCSD(T) are virtually identical, and both methods reproduce the reference interaction energies. The correlation coefficient \(r^2\) with respect to the reference data is 0.996 for both methods, denoting a very good performance to describe the trends in interaction energies of these systems. 
The mean absolute error (MAE) for 
DLPNO-CCSD(T) is 2.1/1.8 kcal/mol for hydrogen-bonded/stacked base pairs, whereas the MAE for LMKLL is only 1.1/2.1 kcal/mol, respectively. In the case of systems where H-bonding links the bases, the interaction energies obtained by LS-DFT are systematically higher in absolute value than those obtained with DLPNO-CCSD(T). In contrast, for the stacked base pair systems, the $\Delta$E\textsubscript{int} are similar in all cases. On the other hand, the LS-DFT H-bonded systems are slightly more accurately described than stacked ones. This fact could be due to the use of the LMKLL functional, which improves the description of non-covalent interactions, but especially when hydrogen bonding plays an important role.\cite{lee2010higher}  
On the other hand, the SE methods considerably underestimate the $\Delta$E\textsubscript{int} for these systems, with  a MAE of 9.2 and 11.6 kcal/mol for PM6-DH2 and PM7, respectively, in the case of H-bonded base pairs and 6.1 kcal/mol and 7.4 kcal/mol, respectively, for stacked base pairs. Nevertheless, all methods show a similar \(r^2\) correlation with respect to the reference data, indicating that even semi-empirical models may describe qualitatively the trends in interaction energies.

\subsubsection{Intercalated Phenanthroline in DNA base pairs (phen/DNA)}
\begin{figure}[h]
\includegraphics[scale=0.75]{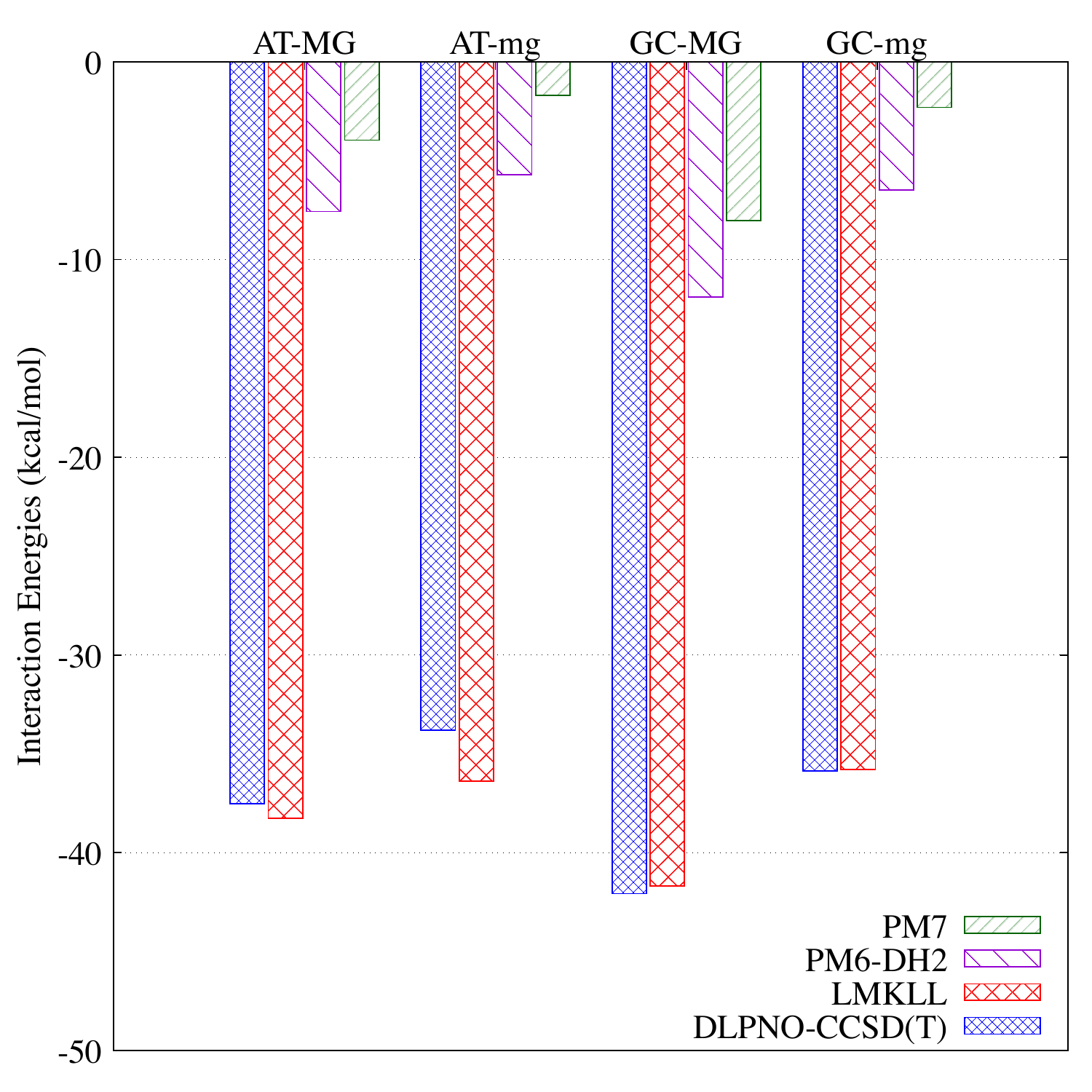}
\caption{Interaction energies (kcal/mol) for the intercalated phenanthroline in DNA base pairs for the DLPNO-CCSD(T), LMKLL, PM6-DH2 and PM7 methods.}
\label{fgr:energy_dimeros} 
\end{figure}

To validate the adequacy of these methods to treat the interaction of ligands intercalated between DNA base pairs through weak interactions, we used a previously characterized system,\cite{gil2015intercalation} where a phenanthroline ligand is intercalated between two pairs of bases through both major groove (MG) and minor groove
(mg). In this previous work, the MP2/6-31G{*}(0.25) theory level was used, based
on the correction performed by Reha et al., which considers a modification of d-polarized basis function\cite{vreha2002intercalators}, leading to a better agreement with CCSD(T) benchmark energies. We performed calculations
for the same systems in this work, taking the DLPNO-CCSD(T) method as a reference. We computed the $\Delta$E\textsubscript{int} between the ligand (phen)
and the DNA base pairs. This $\Delta$E\textsubscript{int} was calculated
by subtracting from the total energy the energy of the separated
fragments: phen ligand\emph{ }and the DNA fragment. The trends obtained for the interaction energies
are presented in Figure \textcolor{black}{\ref{fgr:energy_dimeros}}, whereas the values are depicted in Table S2 of the ESI.

The agreement between LS-DFT and the reference DLPNO-CCSD(T) results is outstanding. The difference in interaction energies is lower than 1 kcal/mol in most of the cases, whereas the MAE is only 0.9 kcal/mol. It must be mentioned that the LMKLL functional has described with greater accuracy the $\Delta$E\textsubscript{int} of the systems containing guanine and cytosine.  Moreover, the LMKLL functional describes the energetic trends very accurately and give more negative interaction energies for the MG structures than for the mg ones.

On the other hand, we observe that the semi-empirical methods show a poor performance with very low interaction energies, namely $\Delta$E\textsubscript{int}, which are around 30 kcal/mol smaller in absolute value than those of DLPNO-CCSD(T). The MAE for both methods is very significant, 29.4 and 33.3 kcal/mol for PM6-DH2 and PM7, respectively. Nevertheless, it must be said that the trends in energetics, in which MG systems are more stable than mg ones, were correctly described by SE methods. For both PM6-DH2 and PM7 methods, the G-C/phen/C-G structure intercalating via MG was described correctly as the most stable system with a $\Delta$E\textsubscript{int} of -11.9 and -8.0 kcal/mol, respectively.

Summarizing, the description of the intercalation of ligands with DNA through weak interactions requires a reliable method to describe non-covalent interactions. Our results point to LS-DFT with the LMKLL functional as a method with an excellent performance in  $\Delta$E\textsubscript{int} for ligand-DNA systems.

\subsubsection{G-quadruplex structures}

The obtained interaction energies for the systems based on G-tetrads according to Eq. \ref{eqn:Celia-eq} are represented graphically in Fig. 5 and their values depicted in Table S5 of the ESI. 

\begin{figure}
\includegraphics[scale=0.6]{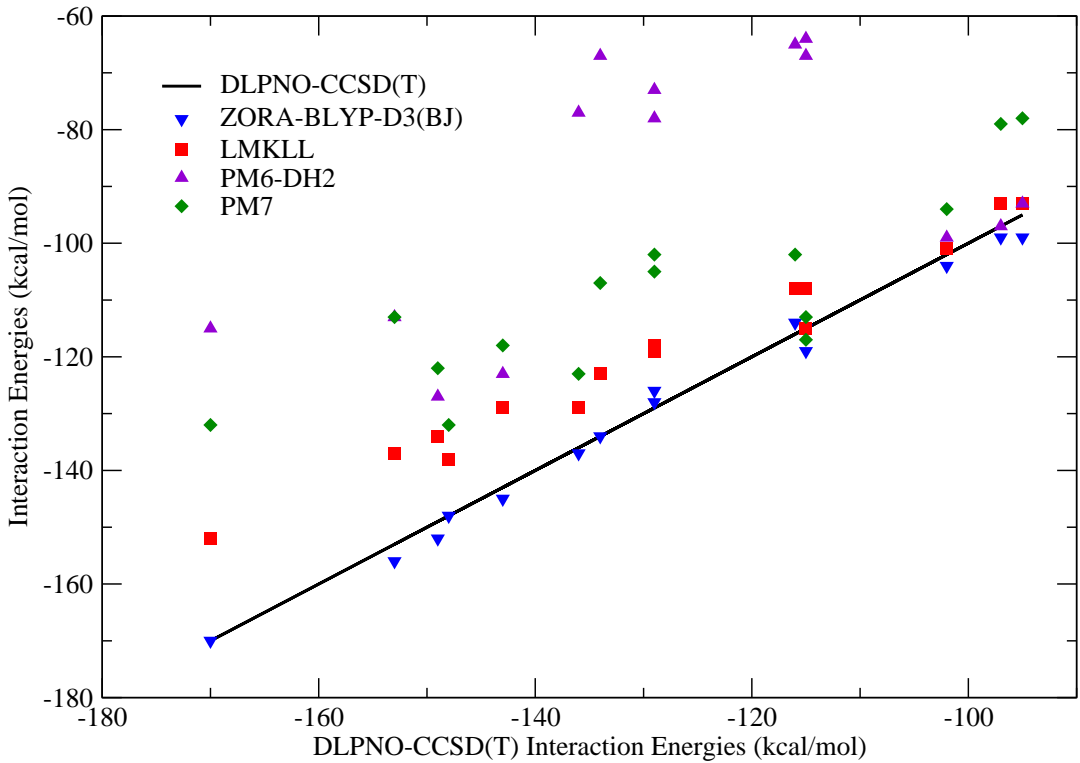}

\caption{Interaction energies (kcal/mol) for the studied structures based on G-tetrads. r\textsuperscript{2} correlation coefficients with respect to the  DLPNO-CCSD(T) reference data for ZORA-BLYP-B3(BJ)\cite{zaccaria2016role} LMKLL, PM6-DH2, and PM7 are 0.993, 0.989, 0.509, 0.777, respectively.}
\label{fgr:energy_GQCelia} 
\end{figure}

LS-DFT agree not only with the DLPNO-CCSD(T) highly correlated benchmark calculations but also with the DFT calculations found in the bibliography for the same systems.\cite{zaccaria2016role} Our LS-DFT results tend to give smaller interaction energies than those found in the literature but the trend is described nicely with a correlation coefficient of 0.989 with respect to the benchmark DLPNO-CCSD(T) calculations. As in the case of the work of Fonseca-Guerra et al.\cite{zaccaria2016role}, the Na metallic cation provides a significant stabilization for the G-tetrad structure and this stabilization decreases as the size of the metallic cation increases, (see Fig. 6). On the other hand, the semi-empirical methods have a very poor behavior both qualitatively and quantitatively. There are significant MAEs for the interaction energies, namely 22.5 and 40.9 kcal/mol for PM7 and PM6-DH2, respectively. Moreover, the poor \(r^2\) correlation coefficients, 0.509 for PM6-DH2 and 0.777 for PM7, reveal high limitations of semi-empirical methods to reproduce the correct trends in stabilization energies. As observed in Fig. 6, the semi-empirical methods show a less clear trend between stabilization of G-tetrads and metal size compared to those observed for DLPNO-CCSD(T) benchmark and LMKLL.

\begin{figure}
\includegraphics[scale=0.6]{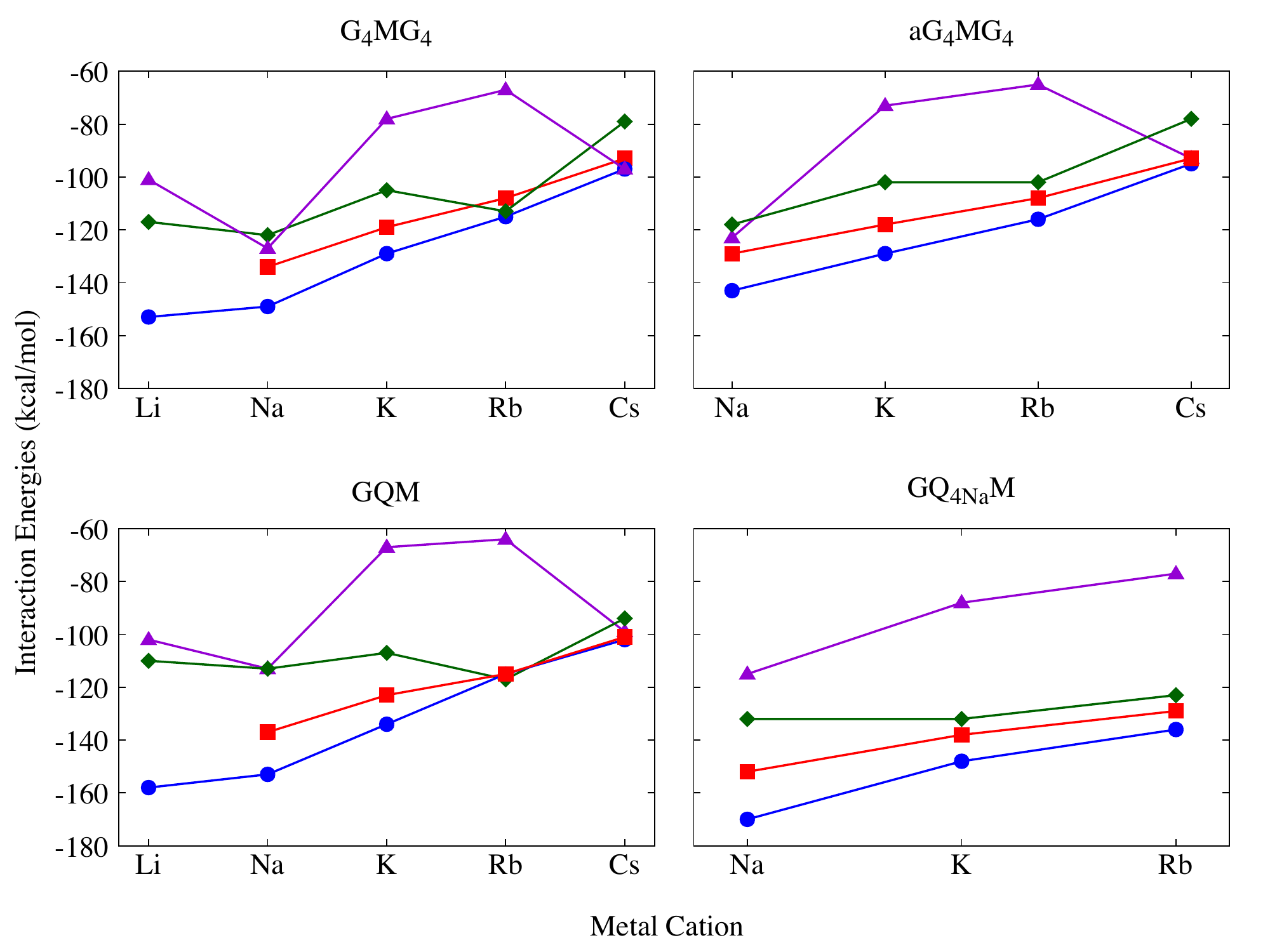}
\caption{Interaction energies (kcal/mol) for the different G-tetrads studied systems (G\textsubscript{4}MG\textsubscript{4} and aG\textsubscript{4}MG\textsubscript{4} on top, and GQM and GQ\textsubscript{4}NaM at bottom) with metal cations (Li, Na, K, Rb, and Cs) for each computational method used in this work. PM6-DH2 in purple, PM7 in green, LMKLL  in red, and DLPNO-CCSD(T) in blue.}
\label{fgr:test_multiplot} 
\end{figure}

To sum up, considering all the interaction energies obtained for the different DNA systems, we can conclude that LS-DFT with the LMKLL functional can describe the weak interactions present in DNA systems. For G-tetrads, the $\Delta$E\textsubscript{int} deviate slightly from the reference ones, but the energetic trends are in good agreement with the reference calculations. However, semi-empirical methods are very limited in the description of interaction energies. Therefore, the use of LS-DFT with the LMKLL functional including van der Waals corrections can be a very suitable strategy to analyze this kind of biological systems, in which a delicate balance of different non-covalent weak interactions is found.

\section{Conclusions}

Over the years, the computational power, the capacity for parallelization, and improvements in software have allowed the study of large biological systems using QM methods. In this sense, approaches based on LS-DFT are 
gaining in speed and efficiency, and they allow a more accurate description of the electronic structure of large biological systems.  In this work
we have studied three different DNA systems: i) DNA base-pairing, ii) models of duplex DNA interacting with a phen ligand and iii) G-tetrads stabilized with various alkaline metals. We have used a LS-DFT method (LMKLL/DZDP) and we have compared its performance with two semi-empirical methods incorporating dispersion corrections, PM6-DH2 and PM7, and QM/MM methods at B3LYP-D3(GD3BJ)/6-31+G(d,p):AMBER and at M11L/6-31+G(d,p):AMBER levels of theory. We have shown how the LMKLL applied through SIESTA reliably predicts both the geometries and the interaction energies for all these DNA systems, using experimental values and DLPNO-CCSD(T) benchmark calculations. On the other hand, the PM6-DH2 semi-empirical method has correctly described the geometries of these systems but not the PM7. However, both semi-empirical approaches are very limited in describing the interaction energies of these systems, with a degree of performance that is system-dependent. The present work opens the door for the computational investigation of large DNA systems using LS-DFT methods, which is particularly interesting for GQ, in which a proper balance of different weak non-covalent interactions, metal-ligand interactions, polarization and charge transfer is needed to give a realistic description of the system. In addition, it must be said that after the consolidation of conventional DFT approaches, different works were reported in which a G2 modified composite methodology where the MP2 geometries and HF frequencies were substituted by the DFT ones and the QCISD(T) computations were replaced by CCSD(T) ones.\cite{mabel1995G2mod} \cite{bauschlicher1995G2mod} We already carried out similar computations by taking into account single-point calculations by means of CCSD(T) on DFT optimized geometries along with thermodynamic corrections to energy also at DFT level with systems of few tens of atoms.\cite{gil2003AHn} \cite{gil2006glygly} \cite{gil2007sidechain} \cite{gil2009ramachandran} In the forthcoming years, we believe that because of the developments in innovative algorithms, software, and hardware, some kind of LS-composite methodology could be feasible by means of single-point near LS-CCSD(T) energies on LS-DFT optimized structures by taking into account biological and chemical systems with thousands of atoms.

\begin{acknowledgement}
This research was financially supported by Diputación Foral de Gipuzkoa through the Gipuzkoa Fellows Program to A. G. This research was also financially supported by the Fundação para a Ciência e a Tecnologia (FCT) by means of Projects PTDC/QUI-QFI/29236/2017, UIDB/04046/2020, and UIDP/04046/2020 and by the Spanish Ministry of Economy, Industry and Competitiveness under the Maria de
Maeztu Units of Excellence Program/MDM-2016-0618). A. G. is thankful to ARAID – Fundación Agencia Aragonesa para la Investigación y el Desarrollo for current funding in the frame of ARAID researchers. The authors also thank for technical
and human support provided by IZO-SGI (SGIker) of UPV/EHU and European funding (ERDF and ESF). Financial support comes also from Eusko Jaurlaritza (Basque Government) through the project IT1254-19, from Spanish MICINN through Grant No.
PID2019-107338RB-C61/AEI/10.13039/501100011033, and grant PGC2018-099321-B-100 from the Ministry of Science, Innovation and Universities through the Office of Science Research (MINECO/FEDER). Pseudopotentials and Numerical Atomic Orbitals
Basis Dataset for SIESTA were provided by Simune Atomistics S. L. (www.simuneatomistics.com). S.E. thanks the Ministry of Higher Education and Scientific Research of Tunisia for the funding of intership carried out in
Portugal and Spain. I.O.d.L. is also very grateful to CICnanoGUNE BRTA for his Ph.D. grant.
\end{acknowledgement}
\begin{suppinfo}
Supporting Information includes definitions of the geometrical
parameters for the duplex DNA and G-quadruplexes, and tables and graphics
with the obtained interaction energies for the DNA base pairs, phen/DNA
intercalation interaction and G-tetrads with the different used methods.
\end{suppinfo}
 \bibliographystyle{plain}
\bibliography{bibliography}

\begin{mcitethebibliography}{71}
\bibitem[Lin and Truhlar(2007)Lin, and Truhlar]{lin2007qm}
Lin,~H.; Truhlar,~D.~G. QM/MM: what have we learned, where are we, and where do
  we go from here? \emph{Theoretical Chemistry Accounts} \textbf{2007},
  \emph{117}, 185--199\relax
\mciteBstWouldAddEndPuncttrue
\mciteSetBstMidEndSepPunct{\mcitedefaultmidpunct}
{\mcitedefaultendpunct}{\mcitedefaultseppunct}\relax
\EndOfBibitem
\end{mcitethebibliography}

\pagebreak{}
\renewcommand{\thepage}{S\arabic{page}}
\setcounter{page}{1}
\begin{center}
\textbf{\large{}Supporting Information for}{\large\par}
\par\end{center}

\begin{center}
\textbf{Semi-empirical and Linear-Scaling DFT Methods to Characterize duplex DNA and G-quadruplexes in Presence of Interacting Small Molecules}
\par\end{center}

Iker Ortiz de Luzuriaga\emph{\textsuperscript{ab}}, Sawssen Elleuchi\emph{\textsuperscript{c}}, Khaled Jarraya\emph{\textsuperscript{c}}, Emilio Artacho\emph{\textsuperscript{adef}}, Xabier Lopez\emph{\textsuperscript{bf\textasteriskcentered}}, and Adrià Gil\emph{\textsuperscript{aghi\textasteriskcentered}}

\emph{\textsuperscript{a}}\emph{CICnanoGUNE BRTA, Tolosa Hiribidea
76, E-20018, Donostia - San Sebastian. }

\emph{}\textsuperscript{\emph{b}}\emph{Polimero eta Material
Aurreratuak: Fisika, Kimika eta Teknologia, Kimika Fakultatea, Euskal
Herriko Uniberstitatea, UPV/EHU, 20080 Donostia, Euskadi, Spain. }

\emph{}\textsuperscript{\emph{c}}\emph{Laboratoire de Chimie Inorganique,
LR17ES07, Université de Sfax, Faculté de Sciences de Sfax, 3000 Sfax,
Tunisia. }

\emph{}\textsuperscript{\emph{d}}\emph{Theory of Condensed Matter, Cavendish Laboratory, University of Cambridge, J. J. Thomson Ave., Cambridge CB3 0HE, United Kingdom}

\emph{}\textsuperscript{e}\emph{Ikerbasque, Basque Foundation for Science, 48011 Bilbao, Spain}

\emph{}\textsuperscript{\emph{f}}\emph{Donostia International Physics
Center, 20018 Donostia, Spain.}

\emph{}\textsuperscript{g}\emph{ARAID Foundation, Zaragoza, Spain}

\emph{}\textsuperscript{h}\emph{Departamento de Química Inorgánica, Instituto de Síntesis Química y Catálisis Homogénea (ISQCH) CSIC- Universitad de Zaragoza, C/ Pedro Cerbuna 12, 50009, Zaragoza, Spain.}

\emph{}\textsuperscript{i}\emph{BioISI – Biosystems and Integrative Sciences Institute, Faculdade de Ciências,Universidade de Lisboa, Campo Grande, 1749-016, Lisboa, Portugal}\bigskip{}

This Supporting Information includes definitions of the geometrical
parameters for the duplex DNA and G-quadruplexes, and tables and graphics
with the obtained interaction energies for the DNA base pairs, phen/DNA
intercalation interaction and G-tetrads with with the different used methods.

\pagebreak{}

\setcounter{figure}{0}
\setcounter{table}{0}
\makeatletter 
\renewcommand{\thefigure}{S\arabic{figure}}
\renewcommand{\thetable}{S\arabic{table}}

\textbf{Definitions for the R and twist angle (\textgreek{j}) parameters
for the duplex DNA systems. }

\begin{figure}
\includegraphics[scale=0.4]{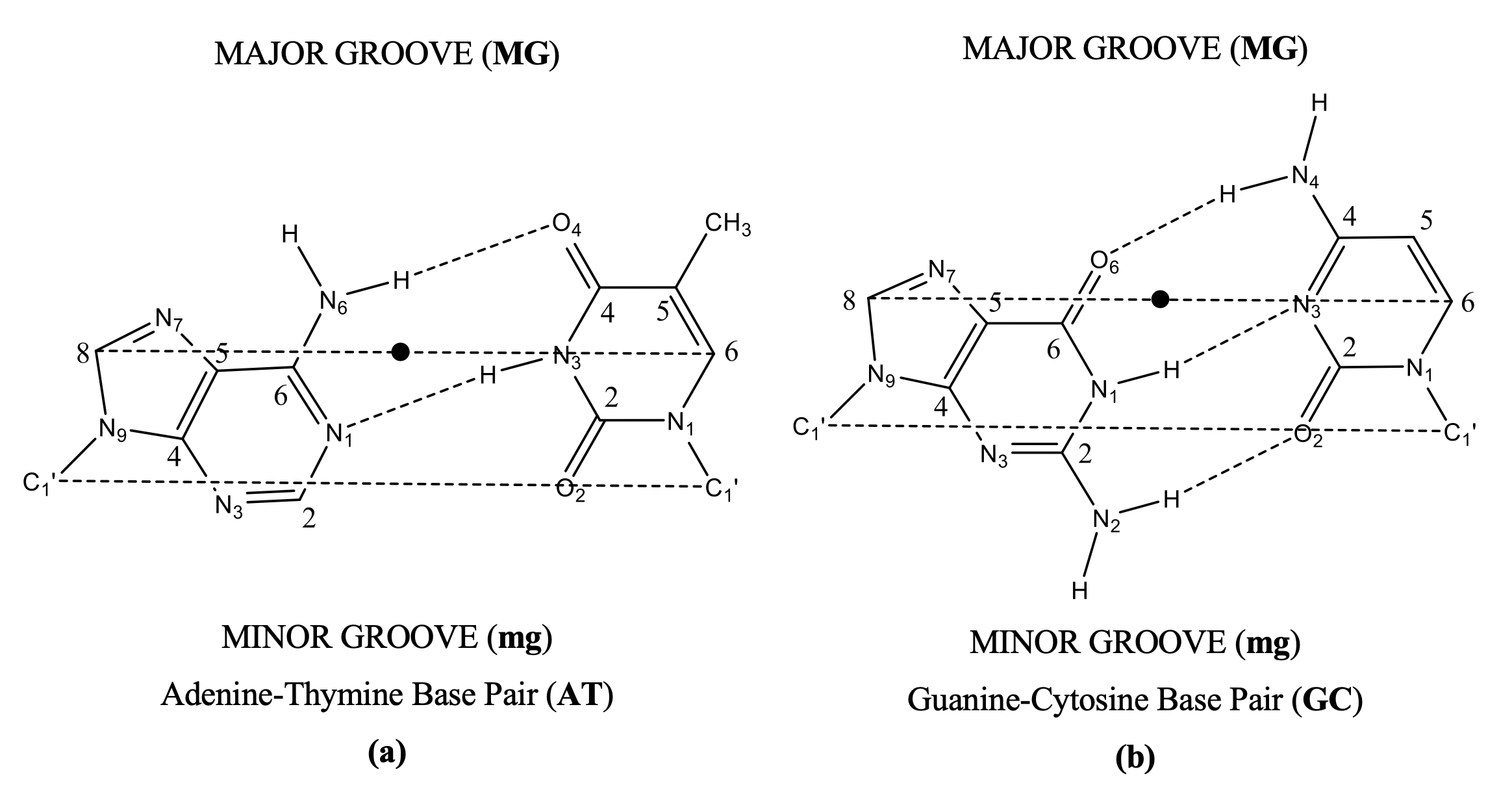}

\caption{Scheme of the base pairs AT (a) and GC (b). The dashed line C\protect\textsubscript{6}--C\protect\textsubscript{8} represents the long base-pair axis, which is roughly parallel to the C'\protect\textsubscript{1}--C'\protect\textsubscript{1} line, where C'\protect\textsubscript{1} stands for the sugar carbon atoms bonded to the bases. The twist angle (\textgreek{j}) is defined as the rotation around the midpoint of the C\protect\textsubscript{6}--C\protect\textsubscript{8} axis (denoted by a dot).}
\label{fgr:Figure_S1} 
\end{figure}

We defined the xy plane by the two atoms forming the N\textsubscript{1}···N\textsubscript{3}
hydrogen bond and the third atom for the definition of the xy plane
is the C\textsubscript{2} atom of adenine (adenine and thymine base pairs) or the C\textsubscript{2} atom of cytosine (guanine and cytosine base pairs), as shown in Figure \ref{fgr:Figure_S1}. Then,
we define the R mean distance between the two base pairs as the difference
between the mean z value of the atoms of the upper base pair and the
one of the atoms of the lower base pair. We also analyzed the
\textgreek{j} twist angle that may be defined from the schemes in Figure \ref{fgr:Figure_S1}. That is, the dashed line
joining the C\textsubscript{8} atom of the purine base to the C\textsubscript{6} atom of the pyrimidine is the
long base-pair axis and the \textgreek{j} angle is defined as the rotation of one
base pair around the center of its C\textsubscript{6}--C\textsubscript{8}
axis. Because the base pairs are not strictly planar and parallel
after optimization, the angle of the optimized systems is defined
as the angle between the projections on the xy plane of the C\textsubscript{6}--C\textsubscript{8}
axis of each base pair. Thus, the \textgreek{j} twist angle would be the dihedral angle between the vector in C\textsubscript{6}--C\textsubscript{8} direction of the i base pair and the vector in the C\textsubscript{6}--C\textsubscript{8} direction of the i+1 base pair in any step of the DNA chain.

\clearpage{}

\textbf{Definitions for the R and twist angle (\textgreek{j}) parameters
for the G-quadruplex systems. }

\begin{figure}
\includegraphics[scale=0.5]{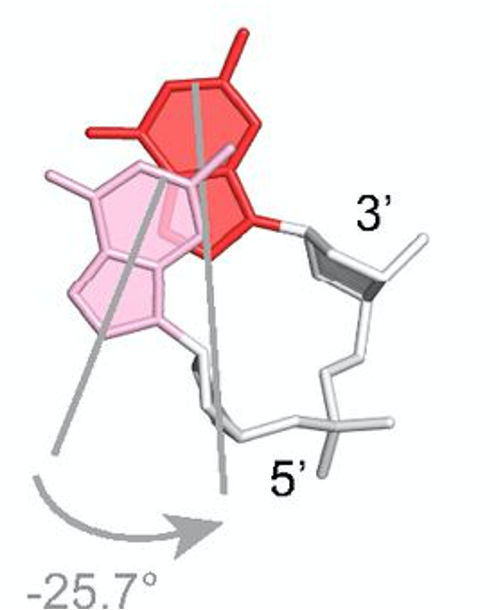}

\caption{Representation of the stacking of two G-tetrad guanines.The \textgreek{j}
twist angle, is defined as the angle between the lines formed by the
guanine C\protect\textsubscript{8} and the midpoint between N\protect\textsubscript{2}
and C\protect\textsubscript{6}.}
\label{fgr:GQ_R_and_twist}
\end{figure}

We defined the xy plane by three guanine O\textsubscript{6} of the
same G-tetrad. Then, we define the R mean distance between the two
G-tetrads as the difference between the mean z value of the atoms
of the upper G-tetrad and the one of the atoms of the lower G-tetrad.
The \textgreek{j} twist angle, is defined as the angle between the
lines formed by the guanine C\textsubscript{8 }and the midpoint between
N\textsubscript{1} and C\textsubscript{2}, as can be seen in Figure \ref{fgr:GQ_R_and_twist}.

\begin{figure}
\includegraphics[scale=0.5]{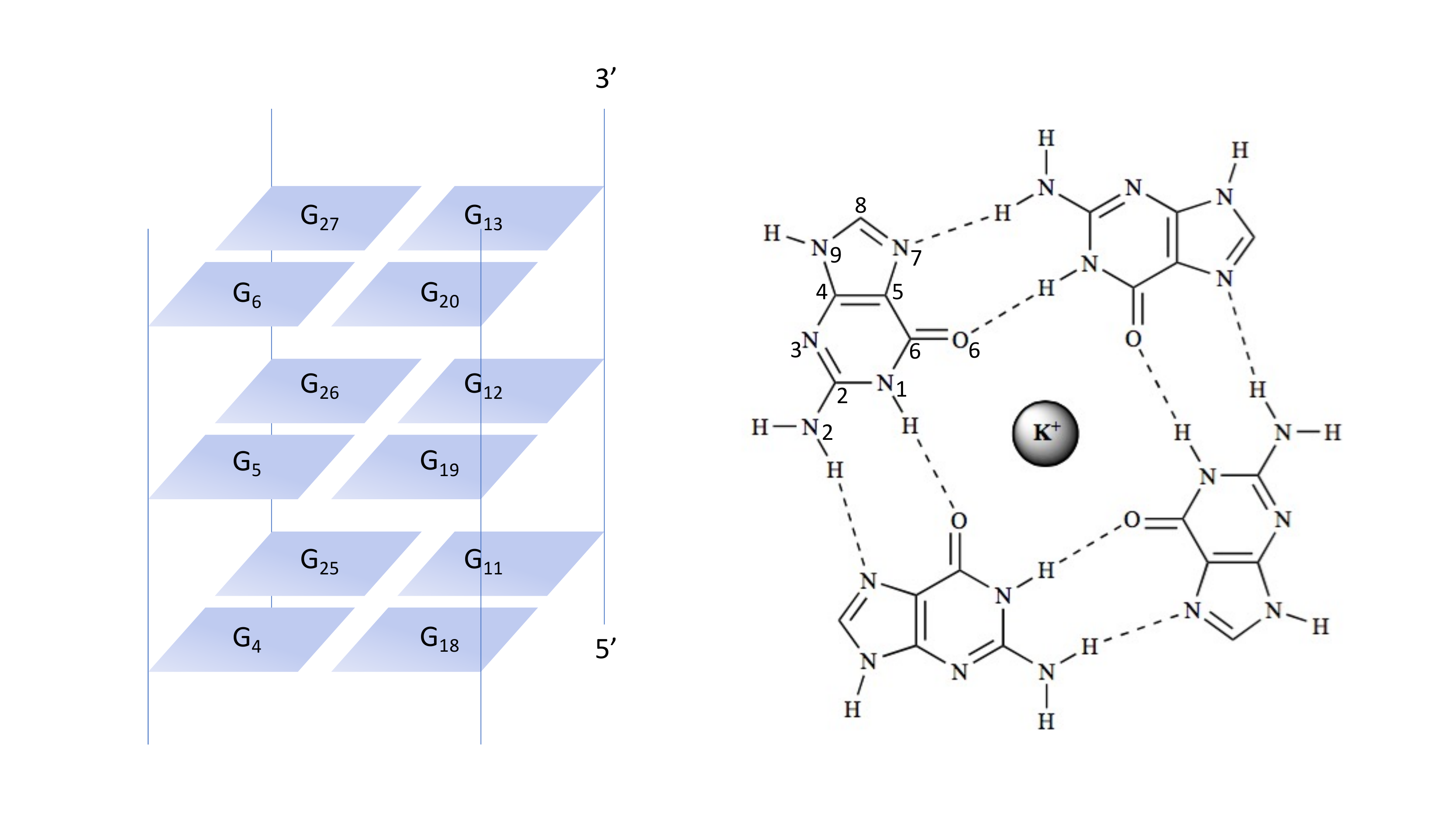}

\caption{On the left, scheme of the stacking of three G-tetrads of guanines composing the 2JWQ PDB structure, and nomenclature for each guanine base. On the right, a G-tetrad, with the number used in the nomenclature.}
\label{fgr:GQ_numeration}
\end{figure}
\newpage{}

\clearpage{}

\begin{table}[H]
\caption{Interaction energies (kcal/mol) for different
DNA base pairs. Abbreviations used in the first column: A, T, C, G
-- adenine, thymine, cytosine, guanine; m -- methyl-; WC, H --
Watson-Crick, Hoogsteen: OG, EG -- optimized geometry, experimental
geometry.}
\label{tbl:Energies_Pavel} %
\begin{tabular}{ccccccc}
& System & Reference & DLPNO-CCSD(T) & LMKLL & PM6-DH2 & PM7\tabularnewline
\hline
\multirow{11}{*}{\rotatebox[origin=c]{90}{H-bonded base pairs}}
&G-C WC (OG) & -32.06 & -33.42  & -32.18  & -18.06  & -16.29\tabularnewline
&mG-mC WC (OG) & -31.59 & -33.53  &  -32.09 & -17.78 &  -16.02\tabularnewline
&A-T WC (OG) & -16.86 & -18.43 & -17.97 & -8.60 & -6.86 \tabularnewline
&mA-mT WC (OG) & -18.16 & -19.54 & -18.84  & -9.09 & -6.26\tabularnewline
&A-T WC (EG)  & -16.40  & -18.32  & -17.95  & -8.46  & -6.81 \tabularnewline
&G-C WC {*} (EG)  & -35.80  & -36.91  & -34.84  & -17.85  & -16.03\tabularnewline
&A-T WC (EG)  & -18.40  & -20.38  & -19.93  & -8.37  & -6.99\tabularnewline
&G-A HB (EG)  & -11.30  & -14.80  & -13.67  & -5.83  & -6.16\tabularnewline
&C-G WC (EG)  & -30.70  & -33.96  & -32.11  & -18.04  & -16.45\tabularnewline
&G-C WC (EG)  & -31.40  & -34.19  & -32.02  & -18.06  & -16.47\tabularnewline
&MAE & -- & 2.08	& 1.08 & 9.15 &	11.61\tabularnewline
\hline
\multirow{13}{*}{\rotatebox[origin=c]{90}{Stacked base pairs}}
&G-C (OG)  & -19.02  & -21.88  & -21.03  & -10.99  & -10.76\tabularnewline
&mG-mC (OG)  & -20.35  & -23.54  & -21.72  & -10.28  & -9.58\tabularnewline
&A-T (OG)  & -12.30  & -14.66  & -14.25  & -5.22  & -4.00 \tabularnewline
&mA-mT (OG)  & -14.57  & -17.52  & -17.37  & -6.14  & -4.65\tabularnewline
&A-T (EG)  & -8.10  & -9.37  & -11.89  & -3.13  & -3.50 \tabularnewline
&G-C (EG)  & -7.90  & -8.13  & -8.84  & -8.00  & -2.90\tabularnewline
&A-C (EG)  & -6.70  & -8.33  & -9.54  & -2.43  & -0.51\tabularnewline
&T-G (EG)  & -6.20  & -7.81  & -9.87  & -3.52  & -3.91 \tabularnewline
&C-G (EG)  & -7.70  & -8.54  & -9.03  & -6.51  & -8.80\tabularnewline
&A-G (EG)  & -6.50  & -9.33  & -8.80  & -6.58  & -4.49\tabularnewline
&C-G (EG)  & -12.40  & -12.71  & -13.11  & -18.06  & -10.44 \tabularnewline
&G-C (EG)  & -11.60  & -12.74  & -12.96  & -10.84  & -16.40\tabularnewline
&MAE  & -- & 1.77 &	2.09 & 6.12 & 7.37\tabularnewline
\hline 
\\
\end{tabular}
\raggedright{}\medskip{}
{*}The geometries of both GC WC (EG) pairs are identical.
\end{table}

\begin{table}[H]
\caption{Interaction energies (kcal/mol) for the stacked
base pairs with the intercalated phen ligand. A-T/phen/T-A MG and A-T/phen/T-A mg corresponds to Adenine-Thymine base pair system with intercalated phen in the major groove (MG) and minor groove (mg), while G-C/phen/C-G MG and G-C/phen/C-G mg corresponds to Guanine-Cytosine base pair system with phen intercalated in the major groove (MG) and minor groove (mg).}

\begin{tabular}{cccccc}
System  & DLPNO-CCSD(T)  & LMKLL  & PM6-DH2  & PM7\tabularnewline
\hline 
A-T/phen/T-A MG  & -37.53  & -38.26 & -7.58 & -3.97\tabularnewline
A-T/phen/T-A mg  & -33.81  & -36.39 & -5.71 & -1.70\tabularnewline
G-C/phen/C-G MG  & -42.06  & -41.69 & -11.91 & -8.03\tabularnewline
G-C/phen/C-G mg  & -35.87  & -35.80 & -6.49 & -2.31\tabularnewline
\hline 
MAE & & 0.94 & 29.40 & 33.32
\end{tabular}

\label{tbl:Energies_dimerosJCTC} 
\end{table}

\begin{table}[H]
\caption{Interaction energies (kcal/mol) for different
DNA base pairs. Abbreviations used in the first column: A, T, C, G
-- adenine, thymine, cytosine, guanine; m -- methyl-; WC, H --
Watson-Crick, Hoogsteen: OG, EG -- optimized geometry, experimental
geometry. The third column correspond to the DFT calculations with
the optimized pseudopotential and basis set. For the fourth and fifth
column psml pseudopotential have been used but, in the fourth column
optimized basis sets were used and for the fifth the default basis sets were used. The Mean Absolute Error (MAE) was calculated taking the experimental data as reference values.}
\label{tbl:Energies_Pavel_psml} %
\begin{tabular}{cccccc}
&System  & Reference  & LMKLL  & \multicolumn{2}{c}{LMKLL/psml}\\
      &  &            &        & Opt. Basis  & Def. Basis\tabularnewline
\hline
\multirow{11}{*}{\rotatebox[origin=c]{90}{H-bonded base pairs}} 
&G-C WC (OG)  & -32.06  & -32.18  & -32.18  & -33.77 \tabularnewline
&mG-mC WC (OG)  & -31.59  & -32.09  & -32.21  & -33.75 \tabularnewline
&A-T WC (OG)  & -16.86  & -17.97  & -18.16  & -19.13\tabularnewline
&mA-mT WC (OG)  & -18.16  & -18.84  & -19.08  & -20.56\tabularnewline
&A-T WC (OEG)  & -16.40  & -17.95  & -18.21  & -19.06\tabularnewline
&G-C WC {*} (EG)  & -35.80  & -34.84  & -35.05  & -36.62\tabularnewline
&A-T WC (EG)  & -18.40  & -19.93  & -20.13  & -20.83\tabularnewline
&G-A HB (EG)  & -11.30  & -13.67  & -14.62  & -14.99\tabularnewline
&C-G WC (EG)  & -30.70  & -32.11  & -32.21  & -34.38\tabularnewline
&G-C WC (EG)  & -31.40  & -32.02  & -32.05  & -34.35\tabularnewline
&MAE & -- & 1.08 & 1.27 & 2.48\tabularnewline
\hline 
\multirow{13}{*}{\rotatebox[origin=c]{90}{Stacked base pairs}}
&G-C (OG)  & -19.02  & -21.03  & -21.20  & -23.14 \tabularnewline
&mG-mC (OG)  & -20.35  & -21.72  & -22.04  & -23.74 \tabularnewline
&A-T (OG)  & -12.30  & -14.25  & -14.70  & -16.01\tabularnewline
&mA-mT (OG)  & -14.57  & -17.37  & -17.79  & -19.04\tabularnewline
&A-T (EG)  & -8.10  & -11.89  & -12.18  & -13.46\tabularnewline
&G-C (EG)  & -7.90  & -8.84  & -8.83  & -11.46 \tabularnewline
&A-C (EG)  & -6.70  & -9.54  & -9.90  & -11.05 \tabularnewline
&T-G (EG)  & -6.20  & -9.87  & -10.26  & -11.24\tabularnewline
&C-G (EG)  & -7.70  & -9.03  & -9.31  & -10.49 \tabularnewline
&A-G (EG)  & -6.50  & -8.80  & -9.36  & -10.44 \tabularnewline
&C-G (EG)  & -12.40  & -13.11  & -13.30  & -15.16 \tabularnewline
&G-C (EG)  & -11.60  & -12.96  & -13.25  & -14.70 \tabularnewline
&MAE & -- & 2.09 & 2.40 & 3.88 \tabularnewline
\hline
\\
\multicolumn{5}{l}{*The geometries of both GC WC (EG) pairs are identical.}
\end{tabular}
\end{table}

\begin{figure}
\includegraphics[scale=0.6]{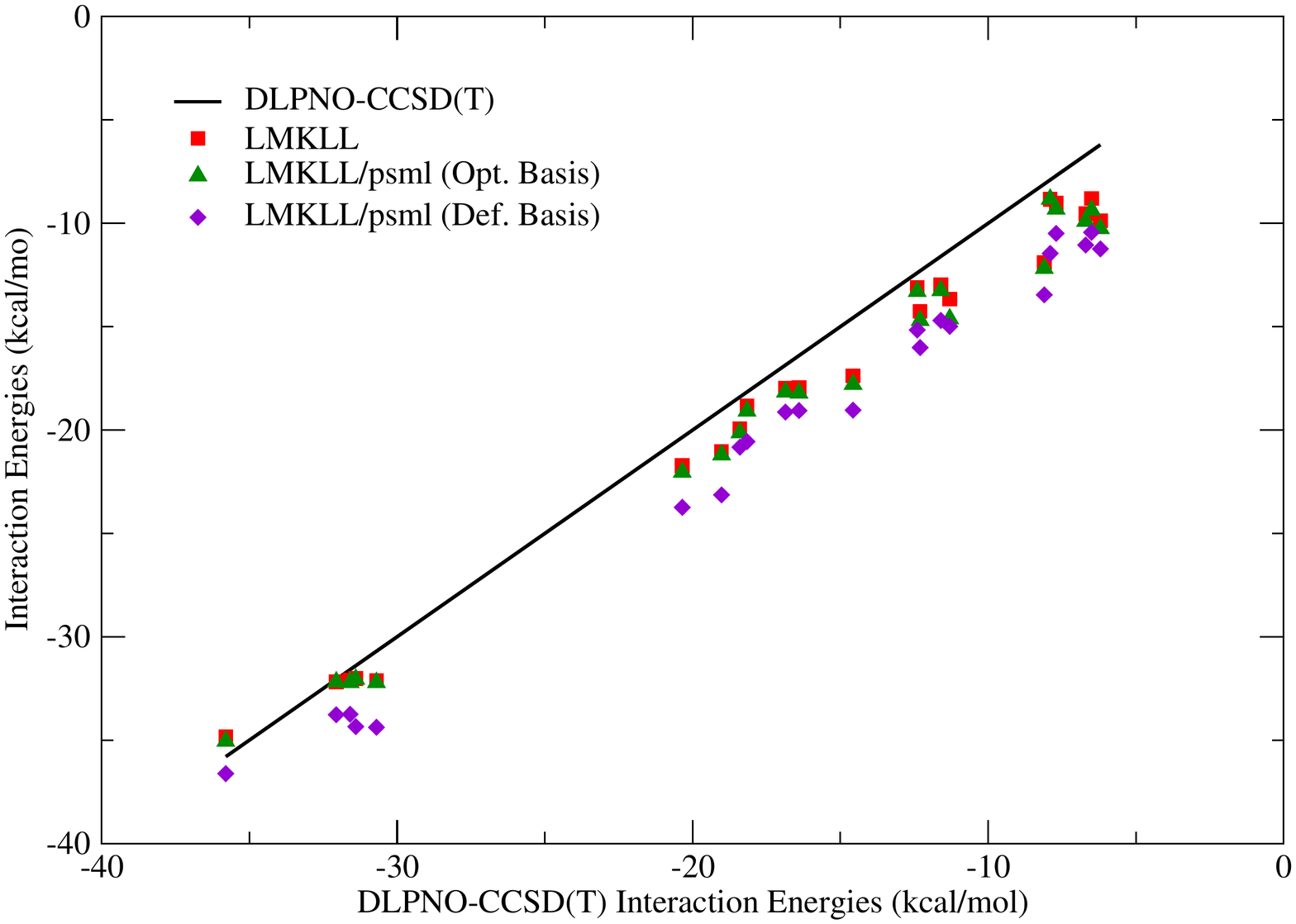}

\caption{Interaction energies (kcal/mol) for the DNA base pair benchmark data set structures\textsuperscript{1} with psml pseudopotentials.
The LMKLL label correspond to the DFT calculations with the optimized
pseudopotential and basis set. In the case of LMKLL/psml (Def. Basis)
the psml pseudopotential were used with the optimized basis sets
and, in the case of LMKLL/psml (Def. Basis) default basis were used. \(r^2\) value for the LMKLL, LMKLL/psml (Def. Basis), and LMKLL/psml (Def. Basis) is 0.997, 0.996, and 0.995, respectively.}

\label{fgr:energy_Pavel_psml} 
\end{figure}

\begin{table}[H]
\caption{Interaction energies (kcal/mol)  of the stacked
base pairs with the intercalated phen ligand. A-T/phen/T-A MG and A-T/phen/T-A mg corresponds to Adenine-Thymine base pair system with intercalated phen in the Major groove (MG) and minor groove (mg), while G-C/phen/C-G MG and G-C/phen/C-G mg corresponds to Guanine-Cytosine base pair system with phen intercalated in the Major groove (MG) and minor groove (mg). The third column corresponds to the LS-DFT calculations
with the optimized pseudopotential and basis set. For the fourth and
fifth column psml pseudopotential have been used but, in the fourth
column optimized basis sets were used and for the fifth the default
basis were used. The Mean Absolute Error (MAE) was calculated taking the DLPNO-CCSD(T) energies as reference values.}

\begin{tabular}{ccccc}

System  & Reference  & LMKLL  & \multicolumn{2}{c}{LMKLL/psml}\\
      &              &        & Opt. Basis  & Def. Basis\tabularnewline
      \hline  
 A-T/phen/T-A MG  &  -37.53  &  -39.17  &  -39.76  &  -49.32 \tabularnewline
 A-T/phen/T-A mg  &  -33.81  &  -36.79  &  -37.59  &  -46.11 \tabularnewline
 G-C/phen/C-G MG  &  -42.06  &  -42.70  &  -43.20  &  -51.89\tabularnewline
 G-C/phen/C-G mg  &  -35.87  &  -36.03  &  -37.33  &  -45.84\tabularnewline
\hline 
 MAE  &  &  0.94  &  2.15  &  10.97
\end{tabular}

\label{tbl:Energies_dimerosJCTC-psml} 
\end{table}

\begin{figure}
\includegraphics[scale=0.75]{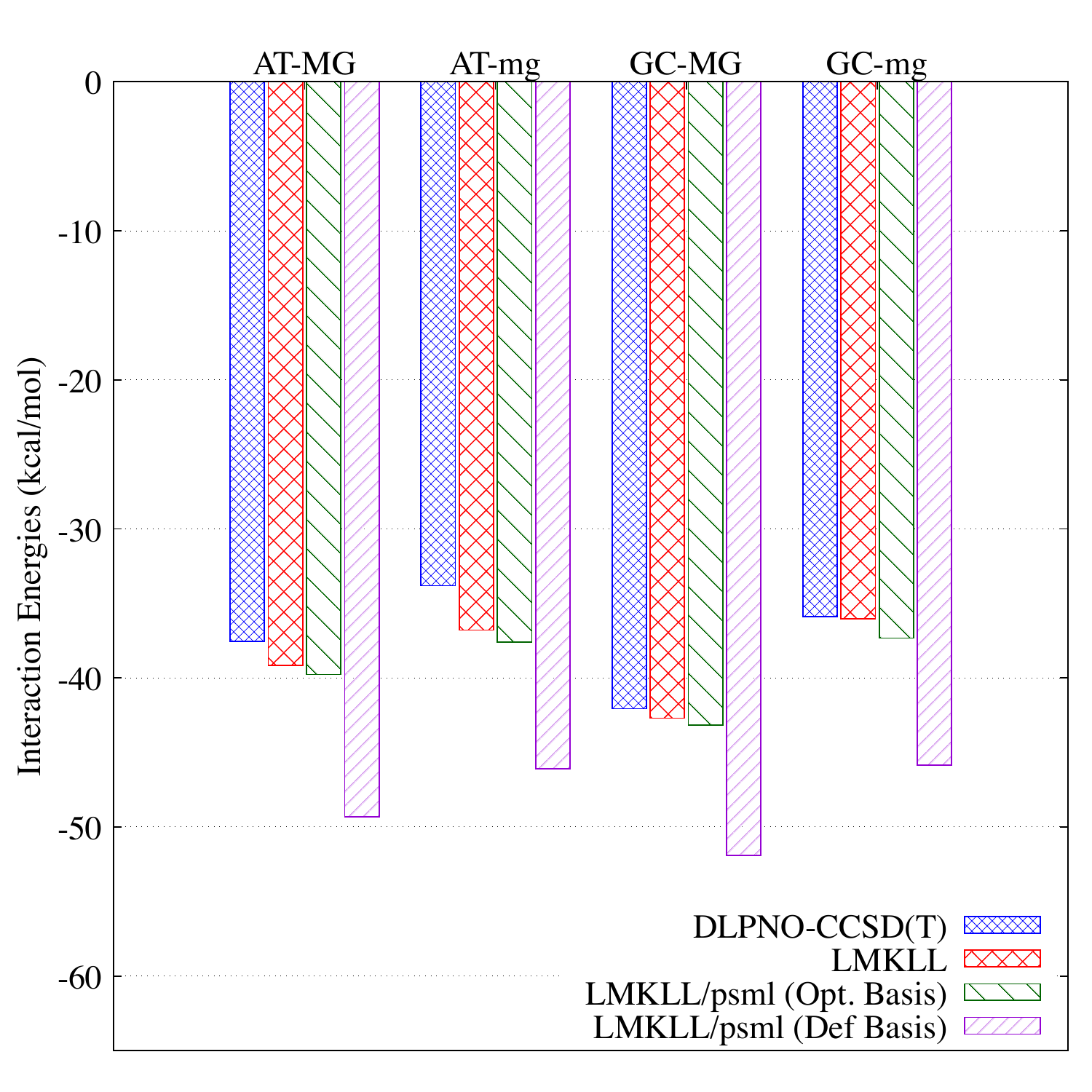}

\caption{Interaction energies (kcal/mol)  for the phen/DNA system with psml pseudopotentials. The LMKLL
label corresponds to the LS-DFT calculations with the optimized pseudopotential
and basis set. In the case of LMKLL/psml (Opt. Basis) the psml pseudopotential
were used with the optimized basis sets and, in the case of LMKLL/psml (Def. Basis) default basis were used.}

\label{fgr:energy_dimerosJCTC_psml} 
\end{figure}

\begin{table}[H]
\caption{Interaction energies (kcal/mol) for the different G-quadruplex structures (G\textsubscript{4}MG\textsubscript{4}, aG\textsubscript{4}MG\textsubscript{4}, GQM, and GQ\textsubscript{4Na}M) with the different metal cations (Li, Na, K, Rb, and Cs) for the used computational methods (PM6-DH2, PM7, LMKLL/DZDP, and DLPNO-CCSD(T)/def2-SVP) along with the results found in the bibliography\textsuperscript{2} at ZORA-BLYP-D3(BJ)/TZ2P level. The Mean Absolute Error (MAE) was calculated taking the DLPNO-CCSD(T)/def2-SVP energies as reference values.}

\begin{tabular}{ccccccc}
 & System  & ZORA-BLYP-D3(BJ) & DLPNO-CCSD(T) & LMKLL & PM6-DH2  & PM7\tabularnewline
\hline 
\multirow{5}{*}{G\textsubscript{4}MG\textsubscript{4}} & Li  & -161.50  & -153.67  & --  & -101.62  & -117.22 \tabularnewline
 & Na  & -152.10  & -149.87  & -134.56  & -127.28  & -122.41 \tabularnewline
 & K  & -128.80  & -129.86  & -119.03  & -78.80  & -105.82\tabularnewline
 & Rb  & -115.50  & -115.54  & -108.90  & -67.96  & -113.17\tabularnewline
 & Cs  & -99.60  & -97.20  & -93.62  & -97.89  & -79.50 \tabularnewline
 \hline 
\multirow{4}{*}{aG\textsubscript{4}MG\textsubscript{4}} & Na  & -145.80  & -143.13  & -129.79  & -123.10  & -118.35\tabularnewline
 & K  & -126.60  & -129.38  & -118.00  & -73.78  & -102.97\tabularnewline
 & Rb  & -114.70  & -116.73  & -108.84  & -65.95  & -102.83 \tabularnewline
 & Cs  & -99.20  & -95.34  & -93.30  & -93.43  & -78.05\tabularnewline
 \hline 
\multirow{5}{*}{GQM} & Li  & -165.70  & -158.36  & --  & -102.40  & -110.31 \tabularnewline
 & Na  & -156.60  & -153.93  & -137.09  & -113.40  & -113.72\tabularnewline
 & K  & -134.70  & -134.41  & -123.74  & -67.00  & -107.12 \tabularnewline
 & Rb  & -119.10  & -115.95  & -115.17  & -64.15  & -117.80 \tabularnewline
 & Cs  & -104.40  & -102.58  & -101.09  & -99.72  & -94.30\tabularnewline
 \hline 
\multirow{3}{*}{GQ\textsubscript{4Na}M} & Na  & -170.90  & -170.04  & -152.29  & -115.60  & -132.87\tabularnewline
 & K  & -148.80  & -148.73  & -138.66  & -88.68  & -132.47\tabularnewline
 & Rb  & -137.30  & -136.45  & -129.67  & -77.25  & -123.23\tabularnewline
\hline 
MAE & & 2.47 & --  & 9.04 & 40.86 & 22.51
\end{tabular}

\label{tbl:Energies_G-quadruplex-Celia} 
\end{table}

\begin{table}[H]
\caption{Total Energy (eV), Wall Time (s) and RMSD (\r{A}) for the G\textsubscript{4}MG\textsubscript{4} system\textsuperscript{2} geometry optimization with different max force tolerance. RMSD value was calculated taking as reference structure the geometry of the literature.}

\begin{tabular}{cccc}

Max Force Tolerance & Total Energy  & Wall Time  & RMSD\\
\hline  
 0.5 & -4659.90  & 19171.3 & 0.01\tabularnewline
 0.2 & -4659.97  & 19287.7 & 0.02\tabularnewline
 0.1 & -4659.99  & 21431.6 & 0.02\tabularnewline
 0.07 & -4660.00 & 23733.8 & 0.03\tabularnewline
 0.05 & -4660.00 & 31654.6 & 0.06\tabularnewline
 0.02 & -4660.01 & 40450.2 & 0.07 \tabularnewline

\end{tabular}

\label{tbl:Test_Tolerance} 
\end{table}

References 
\\
1.- Jurečka, P.; Šponer, J.; Čern`y, J.; Hobza, P. Benchmark database of accurate (MP2 and CCSD (T) complete basis set limit) interaction energies of small model complexes, DNA base pairs, and amino acid pairs. Physical Chemistry Chemical Physics 2006, 8, 1985–1993
\\
2.- Zaccaria, F.; Paragi, G.; Guerra, C. F. The role of alkali metal cations in the stabilization of guanine quadruplexes: why K+ is the best. Physical Chemistry Chemical
Physics 2016, 18, 20895–20904

\end{document}